\documentclass[twocolumn]{aastex631}
\let\tablenum\relax
\usepackage{natbib}
\usepackage{amssymb}
\usepackage{amsmath}
\usepackage{comment}
\usepackage{graphicx}
\usepackage{siunitx}
\usepackage{enumerate}
\usepackage{rotating}
\usepackage{longtable}
\usepackage{booktabs}
\usepackage{float}
\usepackage{array}
\usepackage{gensymb}
\DeclareUnicodeCharacter{03C3}{$\sigma$}
\usepackage{hyperref}
\hypersetup{colorlinks,urlcolor={blue},linkcolor={blue},citecolor={blue}}
\shorttitle{A search for successful and choked jets in nearby SNe Ic-BL}

\shortauthors{O'Dwyer et al.}

\begin{document}

\title{A search for successful and choked jets in nearby broad-lined Type Ic supernovae}

\author[0009-0007-1842-7028]{ Tanner O'Dwyer}
\affiliation{William H. Miller III Department of Physics and Astronomy, Johns Hopkins University, Baltimore, MD 21210 - 21218, USA.}

\author[0000-0001-8104-3536]{Alessandra~Corsi}
\affiliation{William H. Miller III Department of Physics and Astronomy, Johns Hopkins University, Baltimore, MD 21210 - 21218, USA.}

\author[0000-0002-2898-6532]{Sheng Yang}
\affiliation{Institute for Gravitational Wave Astronomy, Henan Academy of Sciences, Zhengzhou 450046, Henan, China}

\author[0000-0003-3768-7515]{Shreya Anand}
\affiliation{Kavli Institute for Particle Astrophysics and Cosmology, Stanford University, 452 Lomita Mall, Stanford, CA 94305, USA}
\affiliation{LSST-DA Catalyst Fellow}

\author[0000-0003-1673-970X]{S.\,Bradley~Cenko}
\affiliation{Astrophysics Science Division, NASA Goddard Space Flight Center, 8800 Greenbelt Road, Greenbelt, MD 20771, USA}
\affiliation{Joint Space-Science Institute, University of Maryland, College Park, MD, 20742, USA}

\author[0000-0002-6428-2700]{Gokul P. Srinivasaragavan}
\affiliation{University of Maryland, College Park, MD, 20742, USA}

\author[0000-0002-9017-3567]{Anna Y. Q.~Ho}
\affiliation{Department of Astronomy, Cornell University, Ithaca, NY 14853, USA}

\author[0000-0003-1546-6615]{Jesper Sollerman}
\affiliation{Department of Astronomy, Oskar Klein Centre, Stockholm University, SE-10691 Stockholm, Sweden}

\author[0000-0003-1600-8835]{Bei Zhou}
\affiliation{Theory Division, Fermi National Accelerator Laboratory, Batavia, Illinois 60510, USA}
\affiliation{Kavli Institute for Cosmological Physics, University of Chicago, Chicago, Illinois 60637, USA}

\author[0000-0003-0477-7645]{Arvind~Balasubramanian}
\affiliation{Indian Institute of Astrophysics, Koramangala II Block, Bangalore 560034, India}

\author[0000-0003-1134-0652]{Po-Wen Chang}
\affiliation{Lawrence Berkeley National Laboratory, Berkeley, CA 94720, USA}

\author[0000-0001-7018-2055]{Marc~Kamionkowski}
\affiliation{William H. Miller III Department of Physics and Astronomy, Johns Hopkins University, Baltimore, MD 21210 - 21218, USA}

\author[0000-0001-8472-1996]{Daniel Perley}
\affiliation{Astrophysics Research Institute, Liverpool John Moores University, 146 Brownlow Hill, Liverpool L3 5RF, UK}

\author[0000-0003-2451-5482]{Russ R. Laher}
\affiliation{IPAC, California Institute of Technology, 1200 E. California Blvd, Pasadena, CA 91125, USA}

\author[0000-0002-5358-5642]{Kohta Murase}
\affiliation{Department of Physics; Department of Astronomy \& Astrophysics; Center for Multimessenger Astrophysics, Institute for Gravitation
and the Cosmos, The Pennsylvania State University, University Park, PA 16802, USA}
\affiliation{Center for Gravitational Physics and Quantum Information, Yukawa Institute for Theoretical Physics, Kyoto University, Kyoto 606-8502, Japan}

\author[0000-0002-8532-9395]{Frank J. Masci}
\affiliation{IPAC, California Institute of Technology, 1200 E. California Blvd, Pasadena, CA 91125, USA}

\author[0000-0002-5619-4938]{Mansi M. Kasliwal}
\affil{Division of Physics, Mathematics, and Astronomy, California Institute of Technology, Pasadena, CA 91125, USA}

\author[0000-0003-1227-3738]{Josiah N. Purdum}
\affil{Caltech Optical Observatories, California Institute of Technology, Pasadena, CA 91125, USA}

\author[0000-0002-3168-0139]{Matthew J. Graham}
\affil{California Institute of Technology, Pasadena, CA 91125, USA}

\begin{abstract}
\label{abstract}
The observational link between long gamma-ray bursts (GRBs) and broad-lined stripped-envelope core-collapse supernovae (SNe Ic-BL) is well established. Significant progress has been made in constraining what fraction of SNe Ic-BL may power high- or low-luminosity GRBs when viewed at small off-axis angles. However, the GRB–SN connection still lacks a complete understanding in the broader context of massive-star evolution and explosion physics. Models predict a continuum of outcomes for the fastest ejecta, from choked to ultra-relativistic jets, and observations from radio to X-rays are key to probing these scenarios across a range of viewing angles and velocities. Here, we present results from a coordinated radio-to-X-ray campaign targeting nearby ($z\lesssim0.1$) SNe Ic-BL designed to explore this diversity. With eight new radio-monitored events and updated data for one previously observed SN, we further tighten constraints on the fraction of SNe Ic-BL as relativistic as SN\,1998bw/GRB 980425. We identify SN\,2024rjw as a new radio-loud event likely powered by strong interaction with circumstellar material (CSM), and add evidence supporting a similar interpretation for SN\,2020jqm. We also establish new limits on the properties of radio-emitting ejecta with velocities consistent with cocoons from choked jets, highlighting SN\,2022xxf as a promising cocoon-dominated candidate. These results refine our understanding of the continuum linking ordinary SNe Ic-BL, engine-driven explosions, and GRBs, and contribute to building a sample that will inform future multi-messenger searches for electromagnetic counterparts to high-energy neutrinos.
\end{abstract}


m\section{Introduction}
Stripped-envelope core-collapse supernovae (SNe) of Type Ic (hydrogen- and helium-poor) with broad lines (BL) constitute $\approx 5\%$ of SNe associated with the deaths of massive stars \citep[][]{2020ApJ...904...35P}. SNe Ic-BL are also the only type of SNe securely associated with Gamma-ray Bursts (GRBs). 
The broad lines that characterize SNe Ic-BL optical spectra point to photospheric velocities systematically higher than those measured in ordinary SNe Ic at similar epochs
\citep{2016ApJ...832..108M,2024ApJ...976...71S}.
The kinetic energies from modeling of optical data of SNe Ic-BL have been inferred to be in the range $(4-7)\times10^{51}$\,erg on average, in excess of the $\approx 10^{51}$\,erg inferred in typical SNe Ibc \citep{2019A&A...621A..71T,2024ApJ...976...71S, 2023ApJ...955...71R}. 

The jet-engine model is a compelling scenario invoked to explain the velocity and energy of SNe Ic-BL \citep[e.g.,][]{2012ApJ...750...68L,2017ApJ...834...28N,2022MNRAS.512.3627D,2022MNRAS.517..582E,2023MNRAS.519.1941P}. In this scenario, successful relativistic jets powering long GRBs are rare outcomes of massive core collapses, and are observed in only $\sim 1\%$ of all SNe Ibc \citep{2010Natur.463..513S}. However, jet engines could still play a role in SNe Ic-BL and low-luminosity GRBs. Low-luminosity gamma-ray bursts (LLGRBs) are a class of GRBs characterized by a lower isotropic-equivalent energy range $10^{46}-10^{48} \text{erg\,s}^{-1}$, than other classes of GRBs \citep{Liang_2007, Nakar2015, https://doi.org/10.1155/2017/8929054}. In fact, a more common outcome of the jet-engine model is that of a choked jet---one that is unable to break out of the surrounding dense matter from the stellar atmosphere which the jet has to cross to produce the bright $\gamma$-rays observed in a long GRB. Even though a choked jet does not emerge, a cocoon forms as the jet drives its way through the stellar envelope, spilling hot material sideways. This cocoon may produce observable signatures \citep{2012ApJ...750...68L,2017ApJ...834...28N,2018ApJ...863...32D,2022MNRAS.512.3627D}. Specifically, the breakout of the cocoon can produce a bright flash sufficiently powerful to explain the origin of low-luminosity GRBs, such as the famous GRB\,980425, associated with the SN Ic-BL 1998bw \citep{Patat_2001,1998Natur.395..663K,2012ApJ...747...88N}. Non-thermal synchrotron radiation produced in the cocoon shock front can also explain the radio emission observed in relativistic SNe without a GRB counterpart \citep{2018ApJ...863...32D}, such as SN\,2009bb \citep{2010Natur.463..513S}.

A key question that remains open in the jet-engine scenario is whether SNe Ic-BL that are not associated with GRBs harbor jet engines: if they do then their jets could be choked, or successful but largely off-axis.

SNe Ic-BL are now being discovered (and their discoveries promptly announced publicly) by large time-domain surveys of the sky at a much increased rate, enabling systematic follow-up observations. Observations of the closest SNe Ic-BL with the Karl G. Jansky Very Large Array \citep[VLA;][]{2011ApJ...739L...1P} presented in \citet{2023ApJ...953..179C} and \citet{2024ApJ...976...71S} have established two key facts. First, SNe Ic-BL (as opposed to any SN Ibc) producing 1998bw-like (i.e., low-luminosity GRB) jets are rare \citep[$<19\%$;][]{2023ApJ...953..179C}, though they still may be more common than  those producing long GRBs \citep[$\approx 1-5\%$ of the SNe Ibc;][]{2006ApJ...638..930S}. Second, the population of SNe Ic-BL with radio detections and radio non-detections are indistinct from one another with respect to their optically-inferred explosion properties, and there are no statistically significant correlations present between the events' radio luminosities and optically-inferred explosion properties \citep{2024ApJ...976...71S}. Hence, optical data alone cannot provide inferences on the radio properties of SNe Ic-BL as related to their fastest ejecta, underscoring the importance of radio observations.

\begin{table*}
\begin{center}
\caption{The sample of 9 SNe analyzed in this work. For each SN we list the IAU name, the ZTF name, the MJD of discovery ($T_0$), the sky position (Right Ascension and Declination), the redshift ($z$), the luminosity distance ($d_L$), and the spectral type.  \label{tab:sample}}
\label{tab:disc}
\begin{tabular}{llccccc}
\toprule
SN & ZTF name & T$_{0}$ & RA/Dec (J2000) &  $z$& $d_L$  & Type \\
 & & (MJD) & (hh:mm:ss~~dd:mm:ss) & &(Mpc)\\
\hline
2020jqm & ZTF20aazkjfv &58980.27 & 13:49:18.56 -03:46:10.27 & 0.037 & 164 & Ic-BL \\
2022xzc  & ZTF22abnpsou &59869.53 & 12:01:12.59 $+$22:36:55.23 & 0.027 & 119 & Ic-BL \\
2022crr & ZTF22aabgazg 	&59628.50 & 15:24:49.13 $-$21:23:21.73 & 0.019 & 82.2 & Ic-BL \\
2022xxf & ZTF22abnvurz 	&59870.53 & 11:30:05.94 $+$09:16:57.37 & 0.003 & 13.0 & Ic-BL \\
2023eiw & ZTF19aawhzsh 	&60033.40 & 12:28:46.20 $+$46:31:15.64 & 0.025 & 110  & Ic-BL\\
2023zeu & ZTF18abqtnbk &60287.11 & 01:36:49.25 $+$01:35:05.68 & 0.03 & 132 & Ic-BL \\
2024abup & ZTF24abvtbyt 	 &60636.35 & 01:49:11.32 $-$10:25:27.44 & 0.0058 & 25.1  & Ic-BL\\
2024adml & ZTF24abwsaxu &60650.46 & 10:10:40.48 $-$02:26:05.14 & 0.037 & 164 & Ic-BL\\
\hline
2024rjw & ZTF24aayimjt 	&60525.30 & 21:03:10.11 $+$20:45:02.58 & 0.02 & 87.5  & Ic\\
\bottomrule
\end{tabular}
\end{center}
\end{table*}

\begin{table*}
\caption{{Optical properties of the 9 SNe in our sample. We list the SN name; the MJD of maximum light in the $r$ band; the absolute magnitude at $r$-band peak; the absolute magnitude at $g$-band peak; the estimated explosion time in days since $r$-band maximum; the estimated nickel mass; the characteristic timescale of the bolometric light curve; the photospheric velocity; the ejecta mass; and the kinetic energy of the explosion. We note that the Arnett modeling cannot be used on the double-peaked light curves of SN\,2022xxf and SN\,2022xzc; see \S\ref{sec:optical_properties} for more details. The optical properties of SN\,2020jqm were previously discussed in \citet{2023ApJ...953..179C}. See Sections \S\ref{sec:vphot} and \S\ref{sec:optical_properties} for discussion. All times are reported in rest frame.
\label{tab:opt_data}}}
\centering
\begin{tabular}{lcccllllll}
\toprule
\toprule
SN & T$_{r,\rm max}$ & M$^{\rm peak}_{r}$ & M$^{\rm peak}_{g}$  & ${\rm T}_{\rm 0}$-T$_{r,\rm max}$  & $M_{\rm Ni}$   & $\tau_{m}$  & v$_{\rm ph}$($^{a}$) & $M_{\rm ej}$ & $E_{\rm k}$ \\
 &  (MJD) & (AB mag)  & (AB mag) & (d) & (M$_{\odot}$) & (d) & ($10^4\,km\,s^{-1}$) & (M$_{\odot}$) & ($10^{51}$erg) \\
\midrule[0.5pt]
2020jqm & 58996.27 & $-18.19\pm0.02$ & $-17.28\pm0.03$ & $-17.4_{-1.5}^{+1.5}$ & $0.36_{-0.01}^{+0.01}$ & $20.15_{-0.50}^{+0.55}$ & $1.3\pm0.3$ (-0.5)  & $5.6\pm1.3$ & $5.6\pm2.9$ \\
2022xzc & 60061.48 & $-16.64\pm0.02$ & $-15.93\pm0.08$ & $-224.1_{-32.2}^{+32.2}$ & 
- & $19.18_{-0.15}^{+0.13}$ & - & - & - \\
2022crr & 59637.50 & $-17.94\pm0.03$ & $-17.42\pm0.14$ & $-12.8_{-0.5}^{+0.5}$ & $0.18_{-0.01}^{+0.01}$ & $7.28_{-0.40}^{+0.34}$ & $1.5\pm0.2$ (3.0) & $0.8\pm0.1$ & $1.1\pm0.3$ \\
2022xxf & 59950.53 & $-15.69\pm0.06$ & $-15.25\pm0.02$ & $-80.0_{-0.1}^{+0.1}$ & 
- & $10.17_{-0.20}^{+0.15}$ & $1.5\pm0.2$ (-4.0) & $1.7\pm0.2$ & $2.2\pm0.7$ \\
2023eiw & 60059.20 & $-18.45\pm0.02$ & $-18.41\pm0.08$ & $-28.4_{-0.4}^{+0.4}$ & $0.43_{-0.01}^{+0.01}$ & $20.36_{-0.21}^{+0.49}$ & $1.4\pm0.5$ (-8.0) & $6.3\pm2.3$ & $7.4\pm5.9$ \\
2023zeu & 60295.20 & $-17.82\pm0.12$ & $-17.31\pm0.14$ & $-9.5_{-1.5}^{+1.5}$ & $0.19_{-0.01}^{+0.01}$ & $7.60_{-0.13}^{+0.17}$ & $1.9\pm0.5$ (-6.0) & $1.2\pm0.3$ & $2.5\pm1.5$ \\
2024rjw & 60536.32 & $-17.32\pm0.01$ & $-17.09\pm0.04$ & $-12.2_{-0.1}^{+0.2}$ & $0.11_{-0.01}^{+0.01}$ & $8.65_{-0.21}^{+0.18}$ & $1.7\pm0.5$ (-2.0) & $1.3\pm0.4$ & $2.3\pm1.5$ \\
2024abup & 60651.36 & $-17.05\pm0.01$ & $-16.52\pm0.04$ & $-20.7_{-0.3}^{+0.4}$ & $0.10_{-0.01}^{+0.01}$ & $20.31_{-0.68}^{+0.42}$ & $0.8\pm0.1$ (28.0) & $<3.4\pm0.5$ & $>1.3\pm0.4$ \\
2024adml & 60666.50 & $-18.82\pm0.04$ & $-18.40\pm0.10$ & $-21.8_{-0.4}^{+0.5}$ & $0.41_{-0.01}^{+0.01}$ & $13.81_{-0.65}^{+0.62}$ & $0.8\pm0.6$ (2.0) & $1.5\pm1.2$ & $0.6\pm1.0$ \\
\bottomrule
\multicolumn{9}{l}{$^{a}$ Rest-frame phase days of the spectrum that was used to measure the velocity.}\\

\end{tabular}
\end{table*}

In this paper, we further tighten the limits on the rate of SNe Ic-BL as relativistic as SN 1998bw/GRB 980425, expanding upon the results of \citet{2023ApJ...953..179C}, and we establish new constraints on the properties of radio-emitting ejecta with velocities consistent with cocoons potentially produced by choked jets. Besides their importance for understanding the physics of massive star explosions, choked jets are interesting in the context of multi-messenger astronomy \citep[e.g., ][and references therein]{2024FrASS..1101792C,2024A&A...690A.187Z}, as theoretical models suggest that they may lead to high-energy neutrino emission while the co-produced gamma-rays are absorbed \citep[e.g.,][]{2016PhRvD..93h3003S}. In fact, several searches in high-energy neutrino detectors' data have targeted both GRBs \citep[e.g.,][]{2015ApJ...805L...5A,2016ApJ...824..115A,2022ApJ...939..116A} and potential choked-jet emission associated with core-collapse SNe \citep[e.g.,][]{2012A&A...539A..60A,2023ApJ...949L..12A,2024PhRvD.109j3041C} but, so far, these searches have not specifically targeted large samples of SNe Ic-BL  showing evidence for fast ejecta components in the radio.  Our work contributes to building such a
sample.

Our paper is organized as follows. In \S\ref{sec:discovery} we present the multi-wavelength observations of the SNe in our sample. In \S\ref{sec:modeling} we discuss our analysis of the collected multi-wavelength data.  In \S\ref{sec:multimessenger} we discuss the multi-messenger prospects highlighting High-energy neutrino contributions. In \S\ref{sec:conclusion} we summarize our results and conclude.

\begin{figure*}
\centering
\includegraphics[width=.7\linewidth]{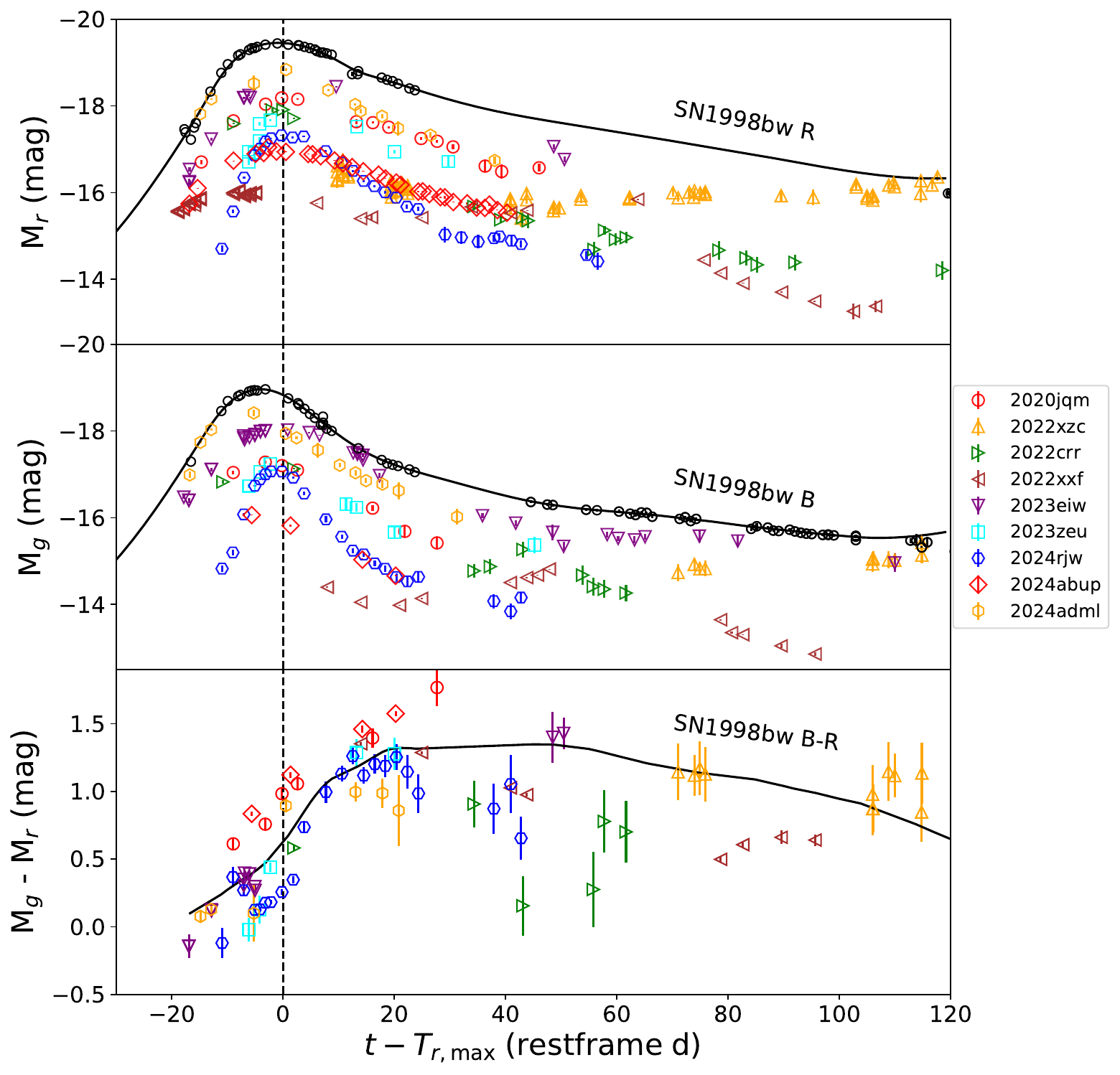}
\caption{The P48 $r$-band (top) and $g$-band (middle) light curves for the SNe in our sample, compared with the R- and B-band light curves of SN 1998bw, respectively. The bottom panel shows the corresponding color evolution, with the archetypal SN\,1998bw represented by \textbf{black open circles}. 
}
\label{fig:1}
\end{figure*}

\begin{figure}
        \centering \includegraphics[width=\columnwidth]{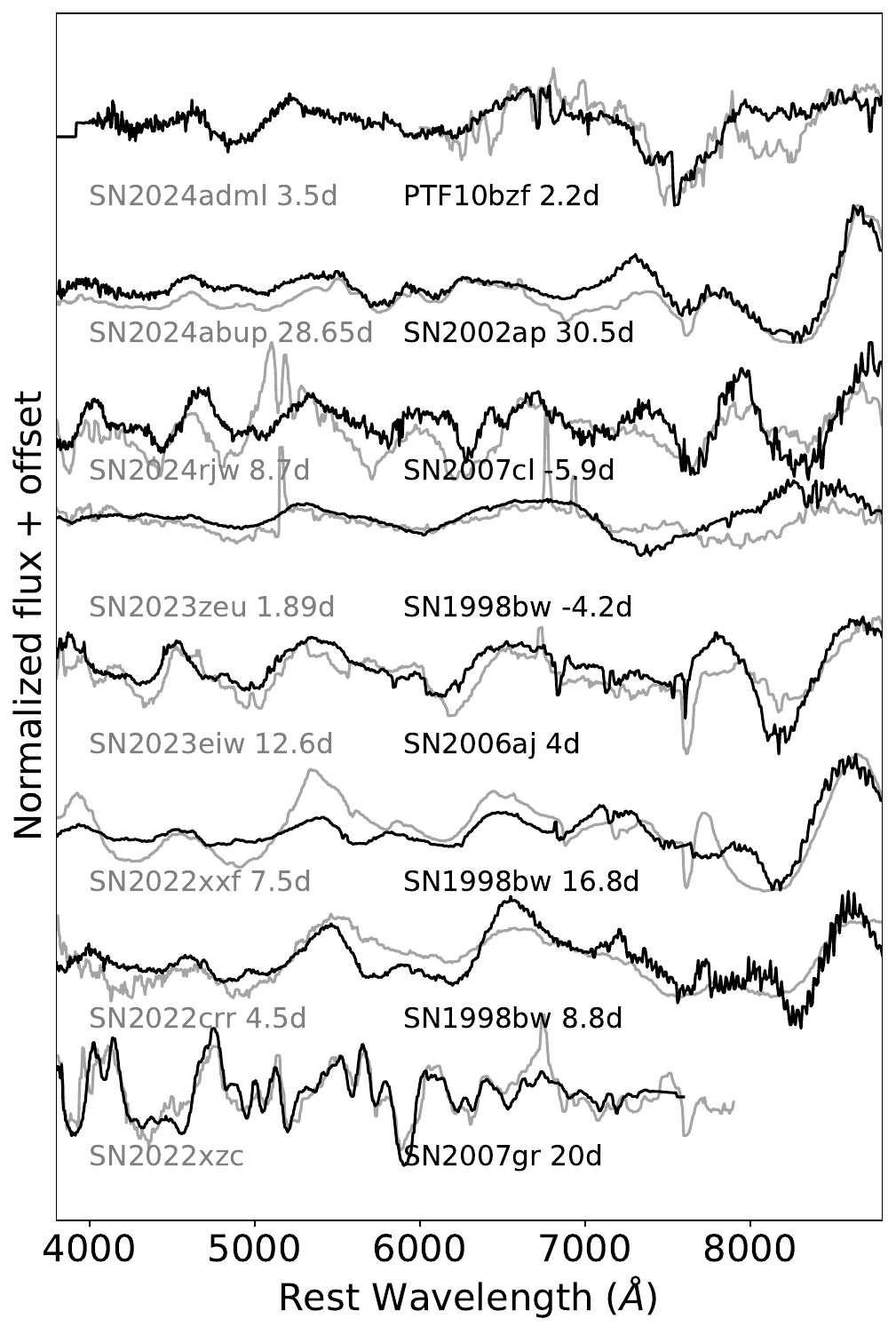}
        \caption{This plot displays the spectra (in gray) along with their best-match templates (in black) from \texttt{Astrodash} for the SNe Ic-BL in our sample. The spectra are labeled with their IAU designation, the IAU name of the best-matched supernova, as well as their phases. Note that for SN\,2022xzc, we do not show the phase since the explosion time is hard to estimate due to its peculiar light curve. The spectra have been selected as the highest resolution ones obtained during the photospheric phase.}
        \label{fig:2}
\end{figure}

\section{Multi-wavelength observations}
\label{sec:discovery}
We have collected a sample of nine supernovae: eight classified as Type Ic-BL and one as Type Ic, observed with the Zwicky Transient Facility \citep[ZTF;][]{Coughlin_2023, 2020PASP..132c8001D, 2019PASP..131a8002B, 2019PASP..131g8001G, 2019PASP..131a8003M, VanderWalt2019} and with follow-up observations in the radio (\autoref{tab:disc}). Of these SNe, 8 were not included in previous radio studies of SNe Ic-BL with deep VLA observations \citep{2023ApJ...953..179C,Corsi-2016}. Two of the SNe (SN\,2020jqm and SN\,2021ywf) were included in \citet{2023ApJ...953..179C}, but here we present additional radio follow-up observations. SN\,2024rjw was initially classified as a Type Ic-BL SN by \cite{2024TNSCR2929....1A}, but later recognized to be a regular SN Ic. We include this source in our sample as it serves as a comparison for nearby stripped-envelope core-collapse SNe that are not expected to host jet engines.

The SNe included in our sample were selected among those monitored through the ZTF bright transient survey \citep[BTS;][]{2020ApJ...895...32F,2020ApJ...904...35P} based on their spectral classification as Type Ic-BL and the availability of VLA observing time. The last was predominantly obtained through joint observing programs of the Neil Gehrels \textit{Swift} Observatory 
\citep{2004ApJ...611.1005G}
and the VLA, targeting SNe Ic-BL located at $d_L\lesssim 150$\,Mpc (this distance enables to probe relativistic ejecta with relatively short VLA observations). 
Given these selection criteria, the median redshift of the SNe in our sample, $z\approx 0.026$, is smaller than that of SNe Ic-BL considered in previous studies such as \citet{2023ApJ...953..179C} (with average redshift $z_{\rm avg}\approx0.037$), \citet{2019A&A...621A..71T} (with average redshift $z_{\rm avg}\approx 0.076$), and of the larger ZTF SN Ic-BL sample presented in \citet{2024ApJ...976...71S} (with average redshift $z_{\rm avg}\approx 0.042$). 

We note that the ZTF has contributed to most of the SN detections considered here. Two of the SNe in our sample (SN\,2022crr and SN\,2024abup) were discovered by the Asteroid Terrestrial-impact Last Alert System \cite[ATLAS;][]{2018PASP..130f4505T}. SN\,2023eiw was first reported by the Panoramic Survey Telescope and Rapid Response System \citep[Pan-STARRS1;][]{2016arXiv161205560C}. In what follows, we describe the observations we carried out for this work. In \S\ref{sec:appendix} we give more details on each of the SNe Ic-BL in our sample.

\subsection{ZTF Photometry}
Photometric observations were obtained with the Palomar Schmidt 48-inch (P48) Samuel Oschin telescope as part of the ZTF survey (\citealp{2019PASP..131a8002B}; \citealp{2019PASP..131g8001G}), using the ZTF camera (\citealt{2020PASP..132c8001D}). In default observing mode, ZTF uses 30\,s exposures, and survey observations are carried out in $r-$ and $g-$ band, down to a typical limiting magnitude of $\approx 20.5$ mag. 
We queried the photometric data from the ZTF forced-photometry service \citep{2019PASP..131a8003M}.   
For SN\,2023eiw and SN\,2024abup, the ZTF observations were sparse, so we used forced photometry data from ATLAS \citep{2018PASP..130f4505T}. 

We corrected all ZTF and ATLAS photometry for Galactic extinction using $E(B-V)$ values toward the SN positions derived from \citet{2011ApJ...737..103S}. All reddening corrections were applied using the \citet{1989ApJ...345..245C} extinction law with $R_V=3.1$. The light curves of all the SNe in our sample are shown in Figure \ref{fig:1}. We discuss the optical light curve analysis in \S\ref{sec:modeling} and summarize the results of this analysis in Table \ref{tab:opt_data}. All the light curves and spectra presented in this work will be made public via Weizmann Interactive Supernova Data Repository (WISeREP; \citealt{2012PASP..124..668Y}).

\subsection{Optical Spectroscopy} 

Preliminary classifications of the SNe in our sample were obtained with either the Spectral Energy Distribution Machine \citep[SEDM; ][]{2018PASP..130c5003B,2019A&A...627A.115R, 2022PASP..134b4505K} or the Double Spectrograph \citep[DBSP;][]{1995PASP..107..375O}. The SEDM is a low-resolution (R $\sim$ 100) integral field unit spectrograph optimized for transient classification with high observing efficiency mounted on the Palomar 60-inch telescope \citep[P60;][]{2006PASP..118.1396C}. The DBSP is a low- to medium-resolution grating spectrograph mounted at the Cassegrain focus of the Palomar 200-inch telescope \citep[P200;][]{1982PASP...94..586O}.

After initial classification, further spectroscopic observations are typically carried out as part of the ZTF transient programs to confirm and/or improve the classification, and to characterize the evolving spectral properties of interesting events. From the series of spectra obtained for each SN, we select one high-quality photospheric-phase spectrum (shown in gray in Figure \ref{fig:2}). We analyze this with \texttt{Astrodash} (\citealt{dash}) to obtain the best match to a SN Ic-BL template (black) after clipping the host emission lines and fixing the redshift to that derived either from the Sloan Digital Sky Survey (SDSS) host galaxy spectrum or from spectral line fitting (H$\alpha$; see the Appendices for further details). In addition to SEDM and DBSP, we also utilized the Low Resolution Imaging Spectrometer (LRIS; \citealt{1982PASP...94..586O}; \citealt{1995PASP..107..375O}), a visible-wavelength imaging and spectroscopy instrument operating at the Cassegrain focus of Keck I \citep{1995PASP..107..375O}; and the Alhambra Faint Object Spectrograph and Camera (ALFOSC), a CCD camera and spectrograph installed at the Nordic Optical Telescope \citep[NOT;][]{2010ASSP...14..211D}.

\begin{table*}
\begin{center}
\caption{\textit{Swift}/XRT observations: SN name; MJD start of the observation; epoch in days between the XRT observation and the estimated SN explosion with error (which is dominated by the uncertainties on the estimated SN explosion time), from explosion epoch which dominate over exposure times; XRT exposure time; 0.3--10\,keV unabsorbed flux and luminosity obtained by co-adding all observations(or $3\sigma$ upper-limit for non detections). }
\label{tab:xrt_summary}
\begin{tabular}{llcccc}
\hline
\hline
SN  & T$_{\rm XRT}$ & T$_{\rm XRT}$-T$_{\rm exp}$ & Exp. & $F_{\rm 0.3-10\,keV}$ & $L_{0.3-10keV}$ \\
 & (MJD)  & (d) & (ks)& ($10^{-14}$\,erg\,cm$^{-2}$\,s$^{-1}$) & ($10^{40}$\,erg\,s$^{-1}$)\\
 \hline
2020jqm & 59002.09 & $23\pm1.5$ & 7.4 & $< 3.3 $ & $<11$ \\
2022xzc & 59903.00 & $34\pm0.1$ & 7.4 & $<4.0$ & $<6.7$ \\
2022crr & 59640.46 & $13\pm0.02$ & 3.3 & $<30$ & $<24$ \\
2022xxf & 59882.11 & $6.2\pm0.2$ & 34 & $<2.2$ & $<4.4\times10^{-2}$ \\
2023eiw & 60061.25 & $29\pm0.4$ & 12 & $<3.2$ & $<4.7$ \\
2023zeu & 60292.12 & $5.2\pm1.7$ & 16 & $<2.4$ & $<5.0$ \\
2024rjw & 60537.07 & $12\pm1.5$ & 11 & $<1.7$ & $<1.6$ \\
2024abup & 60636.80 & $0.45\pm0.4$ & 27 & $<2.5$ & $<1.9\times10^{-1}$ \\
2024adml & 60656.30 & $6.1\pm^{0.3}_{0.2}$ & 27 & $<1.3$ & $<4.1$\\
\hline
\end{tabular}
\end{center}
\end{table*}

\subsection{X-ray follow-up observations}
\label{sec:xray}
We observed the SNe presented in this work (Table \ref{tab:sample}) in X-rays using our \textit{Swift}/XRT \citep{2005SSRv..120..165B} Guest Investigator programs (PI: Corsi). X-ray observations of SN\,2020jqm and SN\,2021ywf were already presented in \citet{2023ApJ...953..179C}. 

We analyze the multiple \textit{Swift}/XRT observations per source; (number of observations per source: SN\,2020jqm 3, SN\,2022xzc 1, SN\,2022crr 2, SN\,2022xxf 10, SN\,2023eiw 4, SN\,2023zeu 4, SN\,2024rjw 6, SN\,2024abup 12, SN\,2024adml 9) using the online XRT tools as described in \cite{2009MNRAS.397.1177E}. For count rate to flux conversion, we adopt a power-law model with photon index $\Gamma = 2$, and correct for Galactic absorption. We do not detect any X-ray emission for any of the SNe in our sample. We summarize our $3\sigma$ upper limits in Table \ref{tab:xrt_summary}, and discuss the implications in \S\ref{sec:X-raymodel}.

\subsection{Radio follow-up observations}
We carried out VLA observations of the SNe in our sample under several observing programs (PI: Corsi). The results are presented in Table \ref{tab:2}. 

The VLA data were calibrated in \texttt{CASA} \citep{2007ASPC..376..127M} using the automated VLA calibration pipeline. Manual inspection was conducted for additional radio frequency interference (RFI) identification and flagging. Calibrated images were formed using \texttt{tclean}. The \texttt{imstat} task was used on the residual images to derive root-mean-square (RMS) noise values in circular regions centered at the optical position of each SN with radius $10\times$ the full-width-at-half-maximum (FWHM) of the nominal synthesized beam. We then used \texttt{imstat} on the clean images to estimate the maximum flux density within a circular region centered on the SN optical position, with a radius equal to the FWHM of the synthesized beam. We report an upper limit (or detection) when the maximum peak flux density in the region is $< 3\sigma$ (or $\ge 3\sigma$). For detections, errors on the measured peak flux densities are the quadrature sum of the noise RMS and a systematic absolute flux calibration error estimated as 5\% of the peak flux density. All detections were also checked for extended versus point-like morphology using the \texttt{imfit} task. Resolved detections are marked as  such in Table \ref{tab:2}. 



\begin{footnotesize}
\begin{longtable*}{ccccccccc}
\caption{VLA observations: SN name; mid MJD of the observation; rest frame days since estimated explosion; observing frequency; flux density or $3\sigma$ upper limit; VLA array configuration; FWHM of the VLA nominal synthesized beam; image RMS; VLA observing project code. \label{tab:2}}\\
\toprule
SN & ${\rm T_{VLA}}$ &  ${\rm \Delta T_{\rm VLA}}$  & $\nu$  & $F_{\nu}$ & Conf. & Nom.Beam  & Image  RMS & Project Code\\
&  (MJD) & (Day) & (GHz) & ($\mu$Jy) &  & (arcsec) & ($\mu$Jy) & \\
\hline
2020jqm &  58997.03 & 17 & 5.6& $167\pm12$ & C  & 3.5 & 8.4 &SG0117$^{\textcolor{blue}{b}}$\\
  &59004.03 & 24 & 5.6 & $293\pm17$ &C  & 3.5 & 8.5 &SG0117$^{\textcolor{blue}{b}}$ \\
  
  & 59028.98 & 49 & 14  & $76\pm11$  & B & 0.42 & 9.8 & 20A-568$^{\textcolor{blue}{b}}$ \\
  &          &       & 5.5 & $206\pm14$ & B & 1.0  & 10 &  \\
  &          &       & 3   & $132\pm13$ & B & 2.1  & 11 &  \\
  
  & 59042.95 & 63 & 14  & $60\pm11$  & B & 0.42 & 10 & 20A-568$^{\textcolor{blue}{b}}$ \\
  &          &       & 5.5 & $197\pm13$ & B & 1.0  & 8.9 &  \\
  &          &       & 3   & $194\pm17$ & B & 2.1  & 14 &  \\
  
  & 59066.09 & 86 & 14  & $58\pm10$  & B & 0.42 & 9.7 & 20A-568$^{\textcolor{blue}{b}}$ \\
  &          &       & 5.5 & $135\pm12$ & B & 1.0  & 9.8 &  \\
  &          &       & 3   & $117\pm13$ & B & 2.1  & 12 &  \\
 
  & 59088.03 & 108 & 14  & $99\pm9.1$  & B & 0.42 & 7.6 & 20A-568$^{\textcolor{blue}{b}}$ \\
  &          &        & 5.5 & $159\pm12$ & B & 1.0  & 8.4 &  \\
  &          &        & 3   & $191\pm20$ & B & 2.1  & 16 &  \\
  
  & 59114.74 & 134 & 14  & $520\pm28$ & B & 0.42 & 9.5 & 20A-568$^{\textcolor{blue}{b}}$ \\
  &          &        & 5.5 & $617\pm32$ & B & 1.0  & 9.8 &  \\
  &          &        & 3   & $393\pm23$ & B & 2.1  & 12 &  \\
  
  & 59240.37 & 260  & 5.5  & $704\pm36$ & A & 0.33 & 7.0 & 20B-149$^{\textcolor{blue}{b}}$ \\
  \hline
  2022xzc & 59902.61 & 33  & 5.5 & $46.6\pm8.9$$^{\textcolor{blue}{a}}$ & C & 3.5 & 8.5 & SI1108$^{\textcolor{blue}{b}}$ \\
        & 59976.30 & 107 & 5.5 & $\lesssim 17$ & B & 1.0 & 5.7 & SI1108$^{\textcolor{blue}{b}}$ \\
\hline
2022crr & 59686.42 & 58 & 6 & $\lesssim 9.6$ & A & 0.33 & 3.2 & 22A-463$^{\textcolor{blue}{c}}$ \\
\hline
2022xxf & 59874.71 & 4.9 & 5.5 & $614\pm32$ & C & 3.5 & 9 & SI1108$^{\textcolor{blue}{b}}$ \\
        & 59902.65 & 33 & 5.5 & $163\pm24$ & C & 3.5 & 11 & SI1108$^{\textcolor{blue}{b}}$ \\
        & 59929.45 & 60 & 5.5 & $83\pm9.7$ & C & 3.5 & 9.7 & SI1108$^{\textcolor{blue}{b}}$ \\
        & 59888.69 & 19 & 15 & $77\pm6.3$ & C & 1.4 & 5.0 & 22B-311$^{\textcolor{blue}{b}}$ \\
&          &       & 9  & $152\pm11$ & C & 2.1 & 7.6 &  \\
&          &       & 5.5 & $261\pm18$ & C & 3.5 & 13 &  \\
        
& 59956.41 & 87 & 15 & $15.0\pm4.8$ & $C\rightarrow B$ & 0.42 & 4.7 & 22B-311$^{\textcolor{blue}{b}}$ \\
&          &       & 9 &  $35\pm7.1$ & $C\rightarrow B$ & 0.60  & 5.2 &  \\
&          &       & 5.5 & $61\pm8.6$ & $C\rightarrow B$ & 1.0 & 8.0 &  \\
        
& 60068.11 & 198 & 15 & $\lesssim 15$ & B & 0.42 & 5.1 & 22B-311$^{\textcolor{blue}{b}}$ \\
&          &        & 9  & $\lesssim 20$ & B & 0.60 & 6.5 &  \\
&          &        & 5.5 & $\lesssim 21$ & B & 1.0 & 7.1 &  \\
        \hline
2023eiw & 60061.10 & 28  & 5.5 & $\lesssim 19$ & B & 1.0 & 6.2 &  SS192066$^{\textcolor{blue}{b}}$ \\
 & 60074.02 & 41 & 5.5 & $\lesssim 18$ & B & 1.0 & 5.9 & SS192066$^{\textcolor{blue}{b}}$ \\
 & 60831.03 & 797 & 5.5 & $\lesssim 20$ & C & 3.5 & 6.7 & SS192066$^{\textcolor{blue}{b}}$\\
\hline
2023zeu & 60313.14 &  26 & 5.5& $49\pm7.9^{\textcolor{blue}{a}}$ & D & 12 & 7.7 & SS192066$^{\textcolor{blue}{b}}$ \\
 & 60839.72 & 553 & 5.5 & $36\pm6.9^{\textcolor{blue}{a}}$ & C & 3.5 & 6.6 & SS192066$^{\textcolor{blue}{b}}$ \\
\hline
2024rjw & 60543.14 & 18 & 5.5 & $430\pm23$ &  B & 1.0 & 7.8 & SS203066$^{\textcolor{blue}{b}}$\\
 & 60579.07 & 54 & 5.5 & $2580\pm130$ & BnA & 0.33 & 10 & SS203066$^{\textcolor{blue}{b}}$\\
  & 60658.79 & 134 & 5.5 & $2408\pm120$ & A & 0.33 & 9.0 & SS203066$^{\textcolor{blue}{b}}$\\
\hline
2024abup & 60658.07 & 22 & 5.5 & $\lesssim 22$ & A & 0.33 & 7.3  & SS203066$^{\textcolor{blue}{b}}$\\
 & 60687.97 & 52 & 5.5 & $\lesssim19 $ & A & 0.33 & 6.2  & SS203066$^{\textcolor{blue}{b}}$\\
 & 60831.56 & 195 & 5.5 & $231\pm15^{\textcolor{blue}{a}}$ & C & 3.5 & 9.6 & SS203066$^{\textcolor{blue}{b}}$\\
\hline
2024adml  & 60659.41 & 9.0 & 5.5 & $\lesssim 19$ & A & 0.33 & 6.2 & SS203066$^{\textcolor{blue}{b}}$\\
 & 60684.34 & 34 & 5.5 & $22\pm6.8^{\textcolor{blue}{a}}$ & A & 0.33 & 6.7 & SS203066$^{\textcolor{blue}{b}}$\\
 & 60831.08 & 181 & 5.5 & $68\pm8.9^{\textcolor{blue}{a}}$ & C & 3.5 & 8.2 & SS203066$^{\textcolor{blue}{b}}$\\
\hline
\multicolumn{9}{l}{$^{a}$ Emission likely dominated by the host galaxy (resolved, marginally resolved, or nuclear; see text for discussion).}\\
\multicolumn{9}{l}{$^{b}$ PI: Corsi.}\\
\multicolumn{9}{l}{$^{c}$ PI: Balasubramanian.}\\
\end{longtable*}
\end{footnotesize}

\section{Multi-wavelength analysis}
\label{sec:modeling}

\begin{figure*}
        \centering
        \includegraphics[width=0.7\textwidth]{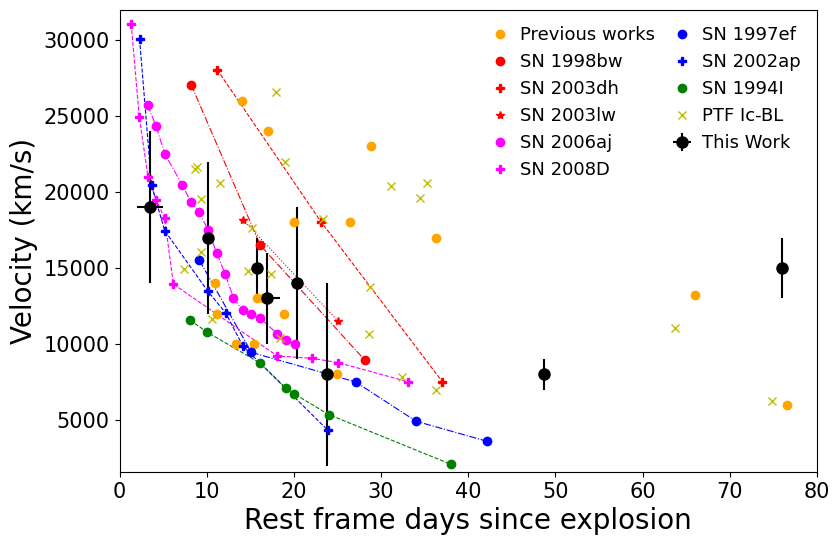}
        \caption{ Photospheric velocities of the 7 Ic-BL and 1 Ic (2024rjw) ZTF SNe with in our sample (black) plotted as a function of time since explosion (see Table \ref{tab:opt_data}). Velocities are measures using Fe II 5169\, \AA\ . We also plot the photospheric velocities of: GRB-SNe \citep[red;][]{1998Natur.395..672I, 2003ApJ...599L..95M, 2006ApJ...645.1323M}; X-ray flash/X-ray transient SNe \citep[magenta;][]{2006Natur.442.1018M, 2006Natur.442.1011P, 2009ApJ...702..226M}; SNe Ic-BL \citep[blue;][]{2000ApJ...545..407M, 2002ApJ...572L..61M}; Type Ic SNe \citep[green;][]{2006MNRAS.369.1939S}. The velocities for the SNe Ic-BL in \citet{Corsi-2016} and \citet{2019A&A...621A..71T} are shown as yellow crosses, those from \citet{2023ApJ...953..179C} and \citet{2024ApJ...976...71S} as orange dots.}
        \label{fig:PhotVel}
\end{figure*}

\subsection{Photospheric velocities}
\label{sec:vphot}
We confirm the Type Ic-BL classification of each object in our sample by measuring their photospheric velocities ($v_{ph}$). SNe Ic-BL are characterized by high expansion velocities which manifest as line broadening in their spectra. A good proxy for the photospheric velocity is that derived from the maximum absorption position of Fe\,\textsc{II}~$\lambda$5169 (e.g., Modjaz et al. \citeyear{2016ApJ...832..108M}). In practice, we estimate the Fe\,\textsc{II} line velocity by comparing it to an average SN Ic spectral template. We caution that estimating this velocity is not easy given the strong line blending. We first preprocessed one high-quality spectrum per object using the IDL routine \texttt{WOMBAT}, then smoothed the spectrum using the python-based routine \texttt{SESNspectraPCA}\footnote{https://github.com/metal-sn/SESNspectraPCA}, and finally ran \texttt{SESNSpectraLib} (Liu et al. \citeyear{2016ApJ...827...90L}; Modjaz et al. \citeyear{2016ApJ...832..108M}) to obtain final velocity estimates. Measured values for the photospheric velocities, their errors at 84\% confidence ($1\sigma$ single-sided), and the  rest-frame phase in days since maximum $r$-band light of the spectra that were used to measure them are also reported in Table \ref{tab:opt_data}.

In Figure \ref{fig:PhotVel}, we show a comparison of the photospheric velocities estimated for the SNe in our sample with those derived from spectroscopic modeling for a number of SNe Ic. Our measured velocities are consistent, within measurement errors, with those from the previous ZTF SN Ic-BL sample \citep{2023ApJ...953..179C} and the PTF/iPTF sample \citep{2019A&A...621A..71T}. 

The photospheric-phase spectrum available for SN\,2024adml (ZTF24abwsaxu) is very low-resolution and poor quality, resulting in higher uncertainty on the velocity estimate. For SN\,2022xzc, we were unable to get estimated velocities as the spectra show features akin to a SN Ic rather than a SN Ic-BL, and \texttt{SESNSpectraLib} is designed only to measure the spectra of SNe Ic-BL. 

\subsection{Bolometric light curve analysis}
\label{sec:optical_properties}

\begin{figure}
    \centering
    \includegraphics[width=\linewidth]{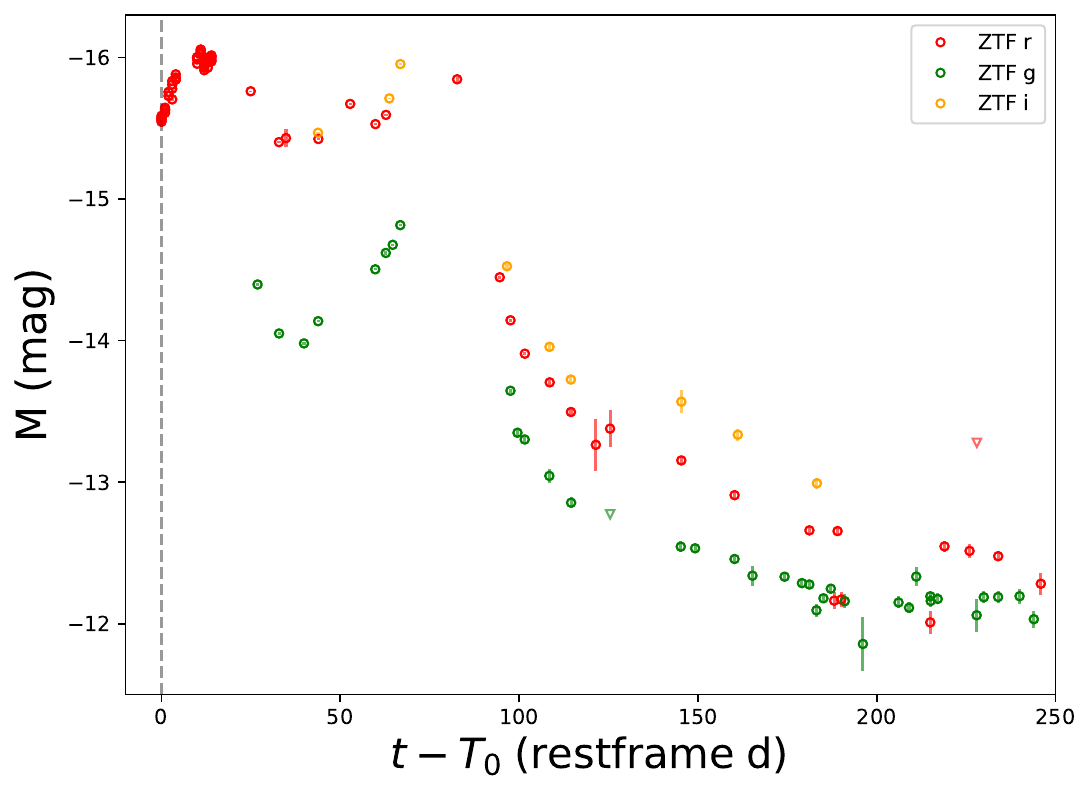}
    \caption{ 
    The $g$-, $r$-, and $i$-band light curves of SN\,2022xxf, with data points shown as circles (green, red, and orange, respectively) and upper limits marked as downward triangles. The light curve clearly displays a double-peaked structure. The vertical dashed line indicates the estimated explosion epoch.
    See \citet{2023A&A...678A.209K} for further a discussion of the interpretation of this double-peaked light curve in the context of ejecta-CSM interaction as a plausible explanation for the second light curve hump. 
    }
    \label{fig:xxf}
\end{figure}

\begin{figure}
    \centering
    \includegraphics[width=\linewidth]{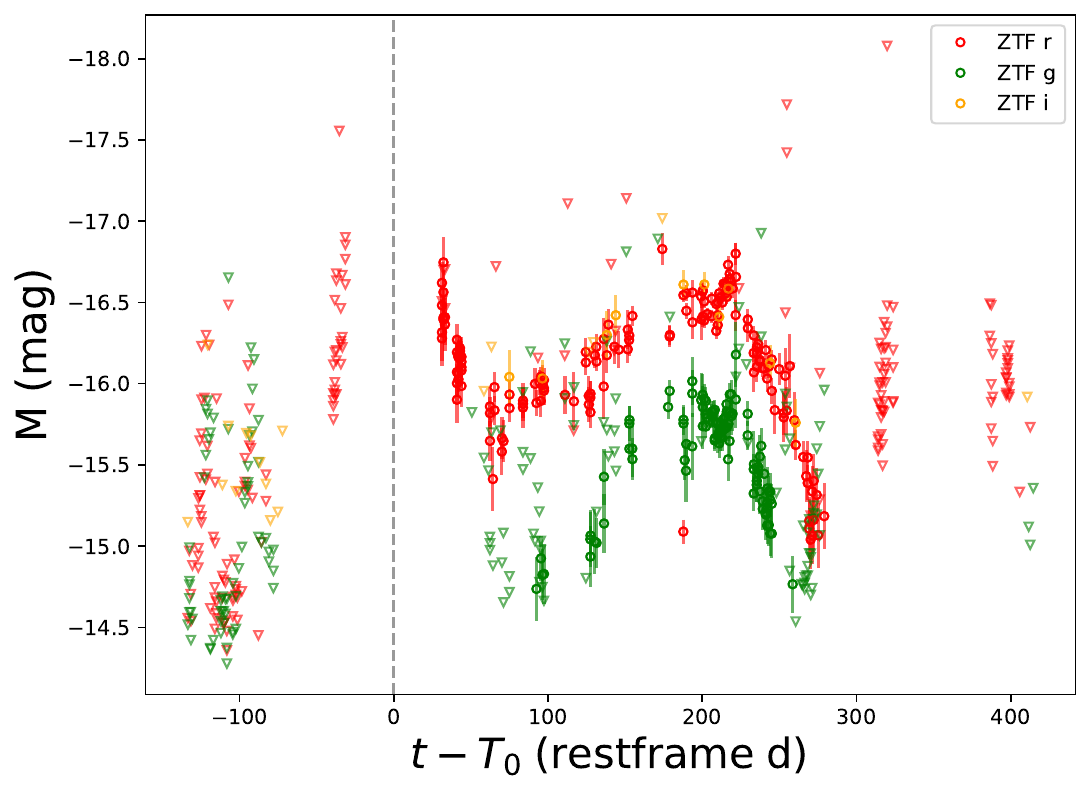}
    \caption{The $g$-, $r$-, and $i$-band light curves for SN\,2022xzc, pointing to a double-peaked structure. 
     }
    \label{fig:xzc}
\end{figure}

\begin{figure*}
    \centering
    \includegraphics[height=8cm]{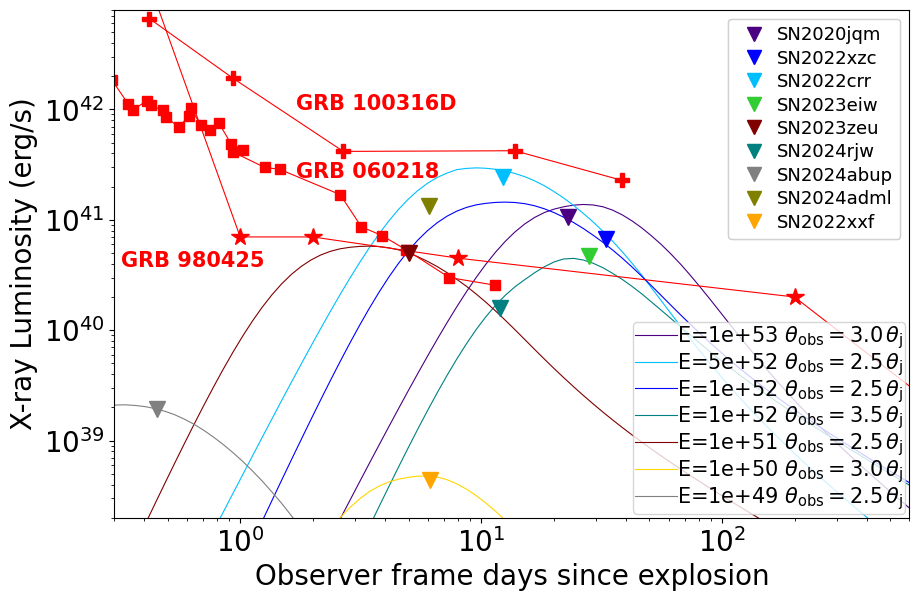}     \caption{\textcolor{magenta}{ } {\it Swift}/XRT upper limits (downward-pointing triangles) for the 8 SNe Ic-BL and for the Type Ic SN\,2024rjw in our sample, compared with the X-ray light curves of three low luminosity GRBs. We can exclude X-ray emission as faint as the afterglow of the low-luminosity GRB\,980425/SN\,1998bw for SN2024rjw, SN2024abup, and SN2022xxf. We also exclude GRB\,060218/SN\,2006aj-like emission for SN\,2022xxf, SN\,2023zeu, SN\,2024rjw, and SN\,2024abup. We can compare our X-ray upper limits with predictions for a range of off-axis GRB afterglow models derived using \texttt{afterglowpy} \citep[dashed lines;][] {2020ApJ...896..166R}. 
      These models assume top-hat jets with isotropic equivalent energies $E$, jet opening angles $\theta_j$, and observers' viewing angles $\theta_{\rm obs}$ as indicated in the legend.  We set $\epsilon_B$ = $\epsilon_e$ = 0.1, and use a constant-density ISM in the range $n = 1–50$\,cm$^{-3}$. All together, our \textit{Swift}/XRT observations rule out GRB X-ray afterglows with kinetic energies $E\gtrsim 10^{51}$\,erg viewed slightly off-axis $\theta_{\rm obs} \approx (2.5-3.5)\theta_j$. For the most nearby events (SN\,2022xxf and SN\,2024abup) our upper-limits also exclude slightly off-axis jets with kinetic energies as low as $\approx 10^{49}-10^{50}$\,erg. }
        \label{fig:X-ray LCs comparison}
\end{figure*}

\begin{figure*}
    \centering
    \includegraphics[width=0.7\textwidth]{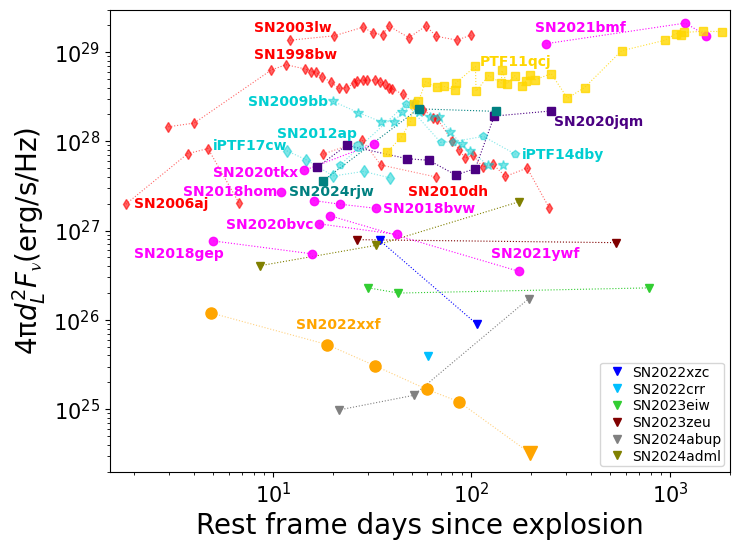}  
    \caption{ Radio ($\approx 6$\,GHz) observations of the SNe Ic-BL in our sample (orange dots and squares for detections, and downward pointing triangles for upper limits; see Table \ref{tab:2}). We compare these observations with the radio light curves of GRB-SNe \citep[red;][]{1998Natur.395..663K,2004Natur.430..648S,2013ApJ...778...18M}, relativistic-to-mildly relativistic SNe Ic-BL discovered independently of a $\gamma$-ray trigger \citep[cyan;][]{2010Natur.463..513S,Corsi-2016,2017ApJ...847...54C,Milisavljevic_2015}, multiple radio detections discussed in the prequel's to this work \citep[magenta;][]{2023ApJ...953..179C,2024ApJ...976...71S, Ho_2019, Ho_2020, Ho_2020_2}, and with PTF11qcj (\citealt{2014ApJ...782...42C}), an example of a radio-loud CSM-interacting SN Ic-BL (yellow).}
    \label{fig:radio_light}
\end{figure*}
We derive the bolometric light curves of the SNe in our sample using the {\tt haffet} framework \citep{Haffet}, which we describe in what follows. First, we correct all ZTF photometry for Galactic extinction, using the Milky Way color excess E(B$-$V)$_{MW}$ toward the positions of the SNe \citep{2011ApJ...737..103S}. All reddening corrections are applied using the \citet{1989ApJ...345..245C} extinction law with $R_V$ = 3.1. We then interpolate our P48 forced-photometry light curves using a Gaussian process (GP) via the \texttt{GEORGE}\footnote{https://george.readthedocs.io/en/latest/} package. 
SNe of the same type are expected to exhibit similar intrinsic spectral energy distribution (SED) evolution in their early phases. As shown in the bottom panel of Figure \ref{fig:1}, all SNe in our sample display very similar early-phase (i.e. 10 rest frame days post maximum light) 
$g-r$ colors as SN 1998bw, suggesting minimal influence from host galaxy extinction. Therefore, we do not account for host galaxy extinction in this study. 
We then calculate ${g}-{r}$ colors using Gaussian process-interpolated light curves and derive bolometric light curves by applying the empirical relations from \citet{2014MNRAS.437.3848L, 2016MNRAS.457..328L}, which provide bolometric corrections based on  $g-r$
color measurements. 

We fit the constructed bolometric light curves with the semi-analytic models developed by \citet{arnett1982}, which relate the light curve properties to the nickel mass ($M_{\rm Ni}$), the characteristic timescale ($\tau_m$), and the time interval between explosion (T$_{exp}$) and peak (e.g. T$_{max}$) epoch (see Table \ref{tab:opt_data}). We note that the Arnett model assumes a SN is powered solely by radioactive decay. Hence, this model may not fully capture the physics of SNe Ic-BL that host a central engine. Nonetheless, previous work \citep[e.g.,][]{2019A&A...621A..71T} has shown that the optical light curves of most SNe Ic-BL are well reproduced by $^{56}$Ni-powered models, with inferred average parameters (M$_{\mathrm{Ni}}$, M$_{\mathrm{ej}}$, and E$_k$) that are, within uncertainties, consistent with those of GRB-SNe. Accordingly, we adopt the Arnett model here, with the caveat that, in the presence of a central engine, the parameter values inferred from the Arnett model should be interpreted as effective descriptors of the light-curve behavior, rather than as a fully accurate representation of the underlying explosion physics.

Because of the sparsity of early-time observations in our sample, power-law fits aimed at determining the time of first light results in poor constraints. Hence, we adopt the Arnett model to estimate the explosion time, using the GP-inferred peak epoch as a reference. This method is applicable primarily to SNe with a single, dominant peak in their light curves. As shown in Figures \ref{fig:xxf} and \ref{fig:xzc}, two sources in our sample (SN\,2022xxf and SN\,2022xzc) exhibit double-peaked light curves \citep[see][]{Sharma2025arXiv250703822S}.  In such cases, the Arnett model fit cannot be used to estimate the nickel masses or explosion times. In fact, when Arnett modeling was attempted on 2022xxf's first peak, the estimated explosion date returned was after the date of optical discovery. Therefore, we set the explosion epoch as the midpoint between the last non-detection prior to discovery, and the first confirmed detection. 

Next, from the measured characteristic timescale $\tau_m$ of the bolometric light curve, and the photospheric velocities estimated via spectral fitting (see \S\ref{sec:vphot}), we derive the ejecta mass (M$_{ej}$) and the kinetic energy (E$_k$) via the following relations \citep[see, e.g., Equations (1) and (2) in][]{2016MNRAS.457..328L}: 

\begin{equation}
    \tau_m^2 v_{\rm ph, max} = \frac{2\kappa}{13.8 c} M_{\rm ej}, \quad v_{\rm ph, max}^2 = \frac{5}{3} \frac{E_k}{M_{\rm ej}},
    \label{eq:1}
\end{equation}
We assume a constant effective optical opacity of $\kappa=0.07$\,g\,cm$^{-2}$. Using $\kappa$ as a constant is justified if electron scattering is the dominant opacity source \citep{1992ApJ...394..599C}. As discussed in \citet{Cano_2013}, various works have adopted values of $\kappa$ in the range $0.05$--$0.08$\,g\,cm$^{-2}$. The ejecta mass and kinetic energy depend on $\kappa$ through the diffusion-time relation (Equation \ref{eq:1}), such that $M_{\rm ej} \propto \kappa^{-1}$ and $E_k \propto M_{\rm ej} v_{\rm ph}^2 \propto \kappa^{-1}$. Hence, the values we quote in Table \ref{tab:opt_data} would be modified by factors of $\sim 0.88-1.4$ over the $0.05$--$0.08$\,g\,cm$^{-2}$ range. 

We note that to derive M$_{ej}$ and $E_k$ as described above we assume the photospheric velocity evolution is negligible within 15 days relative to the peak epoch, and use the spectral velocities measured within this time frame to estimate M$_{ej}$ and $E_k$ in Equation (1) (see Table \ref{tab:opt_data}).

For all SNe in our sample for which we are able to measure photospheric velocities, we report the Table \ref{tab:opt_data} median ejecta masses and kinetic energies of $1.6 \pm 0.6$\,M$_{\odot}$ and $2.3\pm0.8\times 10^{51}$ erg, respectively. These values are most compatible with those of $1.7$\,M$_{\odot}$ and $2.2\times 10^{51}$ erg, reported in \citet{2023ApJ...953..179C}, and close to median values of $1.4$\,M$_{\odot}$ and $2.1\times 10^{51}$ erg, reported in \citet {2024ApJ...976...71S}. Our median values are both a factor of $\approx2$ smaller than those of $3.1$\,M$_{\odot}$ and $5.1\times 10^{51}$ erg, reported in \citet{2019A&A...621A..71T}. We note that SN 2023eiw has the highest $M_{ej}=6.3\pm2.3 \, M_{\odot}$, $E_k=(7.4\pm5.9)\times 10^{51}\,\text{erg}$ which skews the overall mean of sources in Table \ref{tab:opt_data}, without its values the mean of this table becomes comparable to its median at $M_{ej}=1.65~M_{\odot}$, and $E_k=1.67\times 10^{51}$ erg.

\subsection{Search for gamma-rays}
Based on the explosion dates derived in \S\ref{sec:optical_properties} (see also Table~\ref{tab:opt_data}), we searched for coincident GRBs using the Burst Alert Telescope (BAT) aboard the \textit{Neil Gehrels Swift Observatory} \citep{2004ApJ...611.1005G, 2005SSRv..120..143B}, the Gamma-Ray Burst Monitor (GBM) on-board \textit{Fermi} \citep{2009ApJ...702..791M}, and the Konus instrument aboard the NASA Wind spacecraft. We did not include SN\,2022xzc or SN\,2022xxf in the searches, given their uncertain explosion dates (\S\ref{sec:optical_properties}).

No spatial and temporal coincidences were identified with GRBs detected by the BAT or GBM. Several temporal coincidences were found with GRBs detected by Konus-Wind, particularly for SN\,2020jqm, SN\,2023zeu, and SN\,2024abup. However, these are the three events with the least precise explosion date constraints in our sample, and given the search window and the rate of Konus-Wind detections ($\sim0.5$ per day), we expect multiple temporal coincidences by chance. Thus, we cannot robustly associate any of the  SNe Ic-BL in our sample with known GRBs.

\subsection{X-ray constraints}
\label{sec:X-raymodel}
None of the SNe in our sample showed evidence for significant X-ray emission in data collected using  \textit{Swift}/XRT (see \S\ref{sec:xray}). In Table \ref{tab:xrt_summary} we report the 0.3–10\,keV flux upper limits (90\% confidence) derived after correcting for Galactic absorption \citep{2013MNRAS.431..394W}.
In Figure \ref{fig:X-ray LCs comparison} we show these upper limits (downward-pointing triangles) compared with the X-ray observations of GRB-associated SNe. We can exclude X-ray emission as faint as the afterglow of the low-luminosity GRB\,980425/SN\,1998bw for three of the SNe Ic-BL in our sample (SN2024rjw, SN2024abup, and SN2022xxf). Our X-ray observations also exclude GRB\,060218/SN\,2006aj-like emission for SN\,2022xxf, SN\,2023zeu, SN\,2024rjw, and SN\,2024abup. As we discuss in the next Section, while our radio observations can rule out GRB\,980425/SN\,1998bw-like emission for all SNe in our sample (Figure \ref{fig:radio_light}), these X-ray upper limits are unsurpassed in terms of the constraints set on GRB\,060218/SN\,2006aj-like ejecta for the SNe in our sample. This  highlights the value of prompt \textit{Swift}/XRT observations for nearby SNe, especially given the challenges in obtaining prompt VLA follow up. We can also compare our X-ray upper limits with predictions for a range of off-axis GRB afterglow models derived using \texttt{afterglowpy} \citep[dashed lines in Figure \ref{fig:X-ray LCs comparison};][] {2020ApJ...896..166R}. All together, our \textit{Swift}/XRT observations rule out GRB X-ray afterglows with kinetic energies $E\gtrsim10^{51}$\,erg viewed slightly off-axis $\theta_{\rm obs} \approx (2.5-3.5)\theta_j$. For the most nearby events (SN\,2022xxf and SN\,2024abup) our upper-limits also exclude slightly off-axis jets with kinetic energies as low as $\approx 10^{49}-10^{50}$\,erg. 

We note that these constraints depend on the choice of the microphysical parameters $\epsilon_e$ and $\epsilon_B$, with the most critical dependence being on $\epsilon_e$. Specifically, decreasing $\epsilon_e$ by an order of magnitude while keeping the other parameters unchanged would shift the model light curves down by an order of magnitude or more. On the other hand, a change in $\epsilon_B$ has a very small effect. We also note that we assume a constant density interstellar medium (ISM). A stellar wind medium with a density profile $\rho \propto R^{-2}$ would impact off-axis light curve predictions. Generally speaking, for off-axis observers, a wind medium correlates with a shallower rise to the peak of the light curve, and a flatter and wider light curve peak. However, this effect is less pronounced in X-rays (compared e.g., to the optical or radio), since above the cooling synchrotron break, the observed flux density becomes much less sensitive to the external density \citep[see e.g.,][for a discussion]{2018MNRAS.481.2711G,2022Univ....8..588Z}. 

\begin{figure}
        \centering
        \includegraphics[width=1\linewidth]{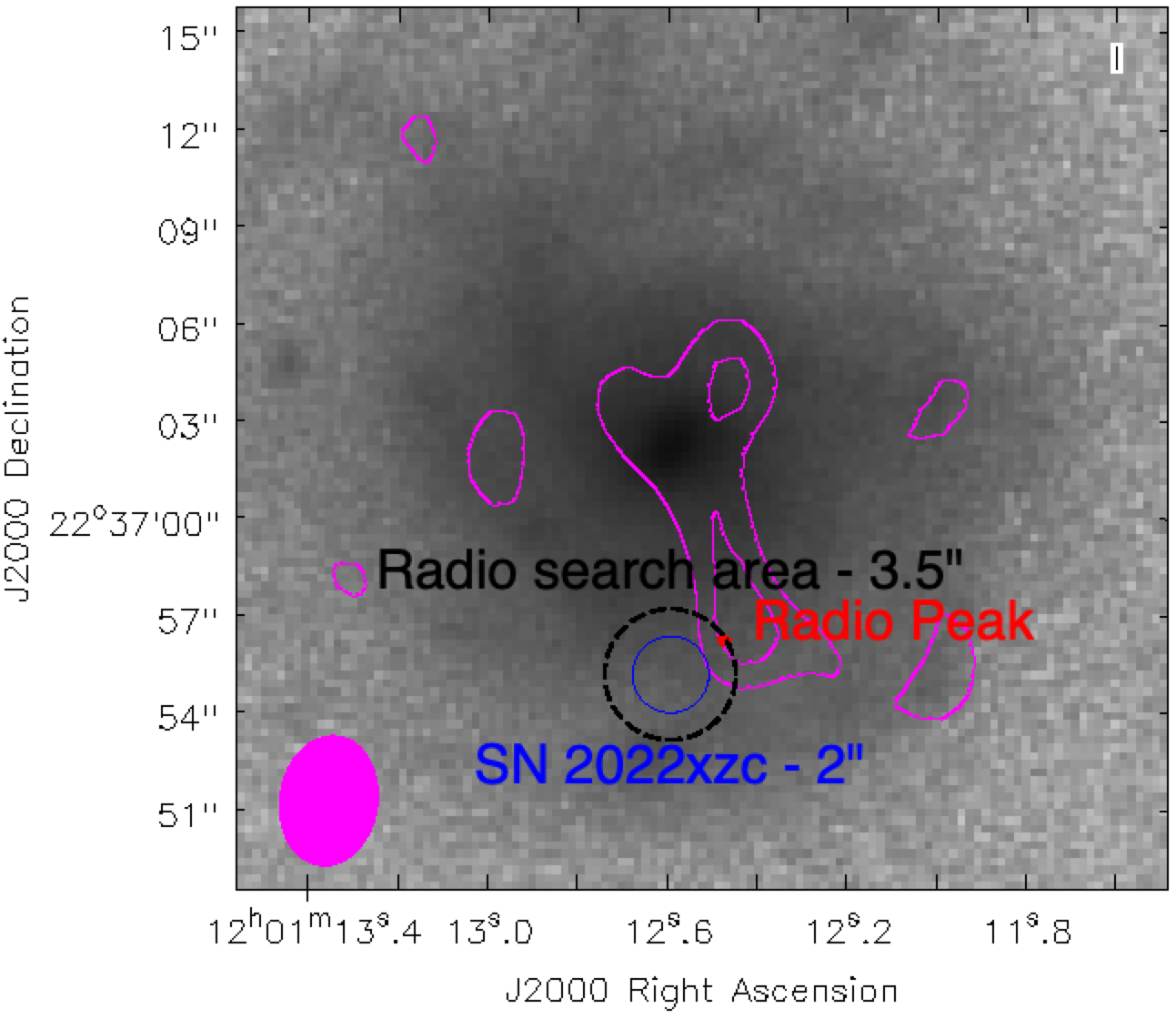}
       \caption{Optical (PanSTARRS-1) image of SN2022xzc with radio contours mapped in magenta. The radio contours correspond to the first VLA epoch taken at $\approx33$ days after the estimated explosion time (at 5.5\,GHz with the VLA in its C configuration). The black circle is centered at the optical SN position and has a radius equal to the nominal VLA synthesized beam FWHM for that epoch (3.5\,arcsec). The blue circle has a radius of 2\,arcsec, comparable with the position accuracy of ZTF. The magenta contour lines show that the majority of the radio emission is centered around the host galaxy rather than close to the optical SN position. The location of the peak (red dot) of the radio excess we measure in our search area (black circle) shows that this emission is likely due to contamination from the host galaxy (see \S\ref{sec:host galaxy emission} for further discussion).}
       \label{fig:xzc emission}
\end{figure}

\begin{figure}
        \centering
        \includegraphics[width=1\linewidth]{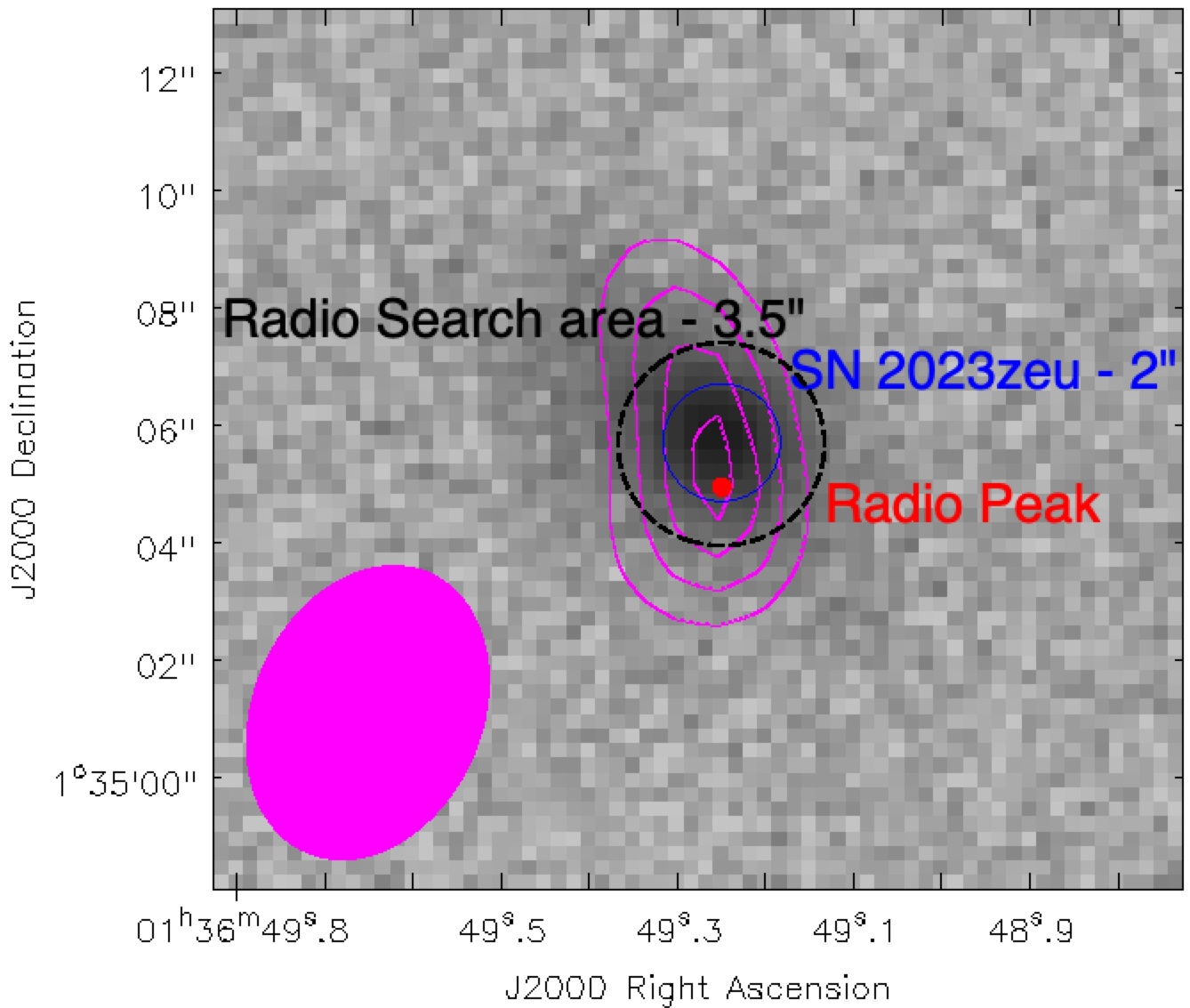}
       \caption{Optical (PanSTARRS-1) image of SN\,2023zeu with radio contours mapped in magenta. The radio contours correspond to the second epoch taken $\approx552$ days after the estimated explosion time (at 5.5 \,GHz with the VLA in its C configuration). The black circle is centered at the optical SN position and has a radius equal to the nominal VLA synthesized beam FWHM for that epoch  (3.5\,arcsec). The blue circle has a radius of 2\,arcsec, comparable with the position accuracy of ZTF. The magenta contour lines show that the majority of the radio emission, including the peak (red dot) of the excess radio emission we measure in our VLA search area (black circle) likely originates from the central region of the host galaxy  (see \S\ref{sec:host galaxy emission} for further discussion).}
       \label{fig:2023zeu}
\end{figure}

\begin{figure}
\centering
    \includegraphics[width=1\linewidth]{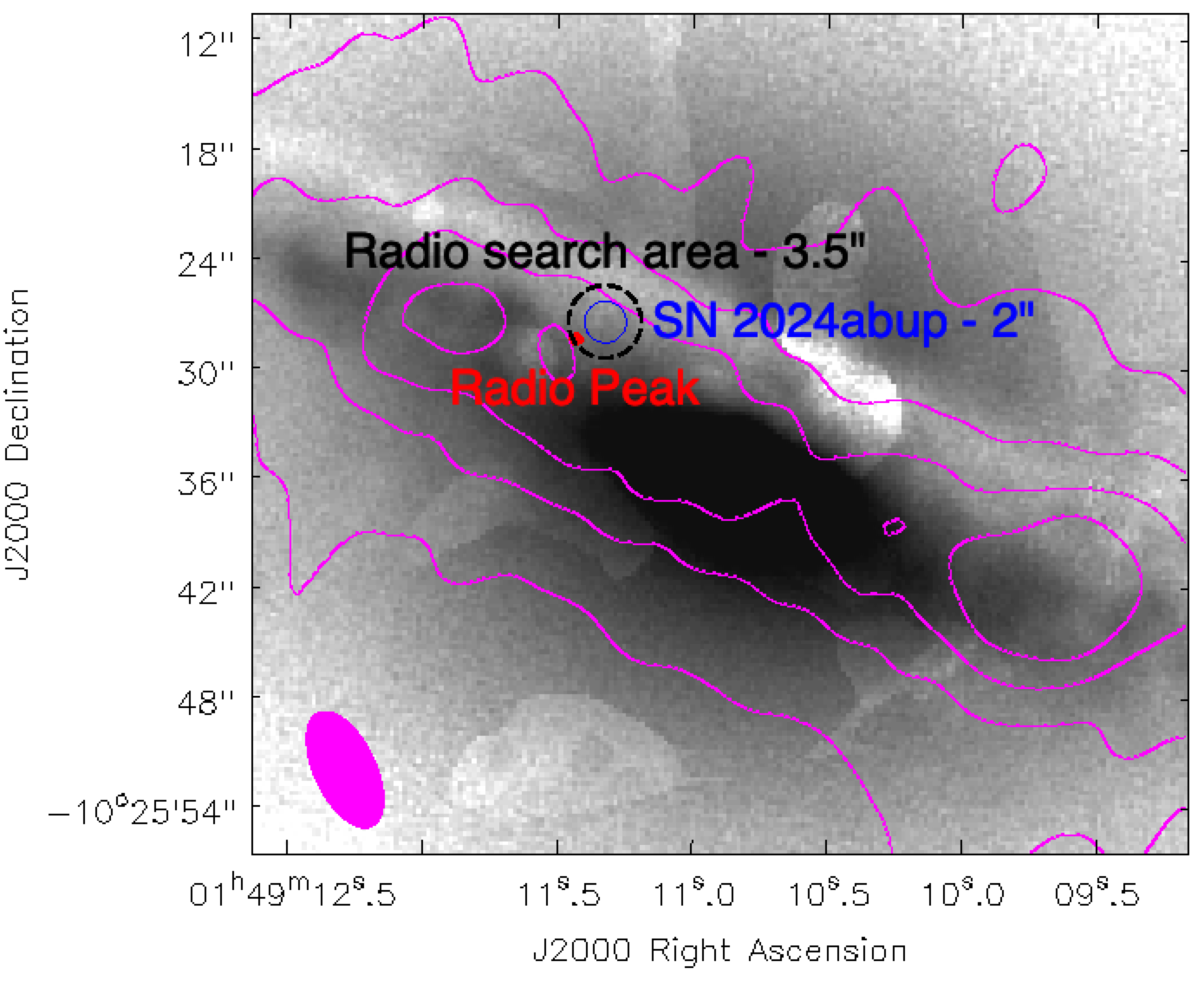}
    \caption{Optical (PanSTARRS-1) image of SN\,2024abup with radio contours mapped in magenta. The radio contours correspond to the third epoch taken $\approx195$ days after the estimated explosion time (at 5.5 \,GHz with the VLA in its C configuration). The black circle is centered at the optical SN position and has a radius equal to the nominal VLA synthesized beam FWHM for that epoch  (3.5\,arcsec). The blue circle has a radius of 2\,arcsec, comparable with the position accuracy of ZTF. The magenta contour lines show that the majority of the radio emission, including the peak (red dot) of the excess radio emission we measure in our VLA search area (black circle) likely originates from the central region of the host galaxy (see \S\ref{sec:host galaxy emission} for further discussion).}
    \label{2024abup}
\end{figure}

\begin{figure}
    \centering
    \includegraphics[width=1\linewidth]{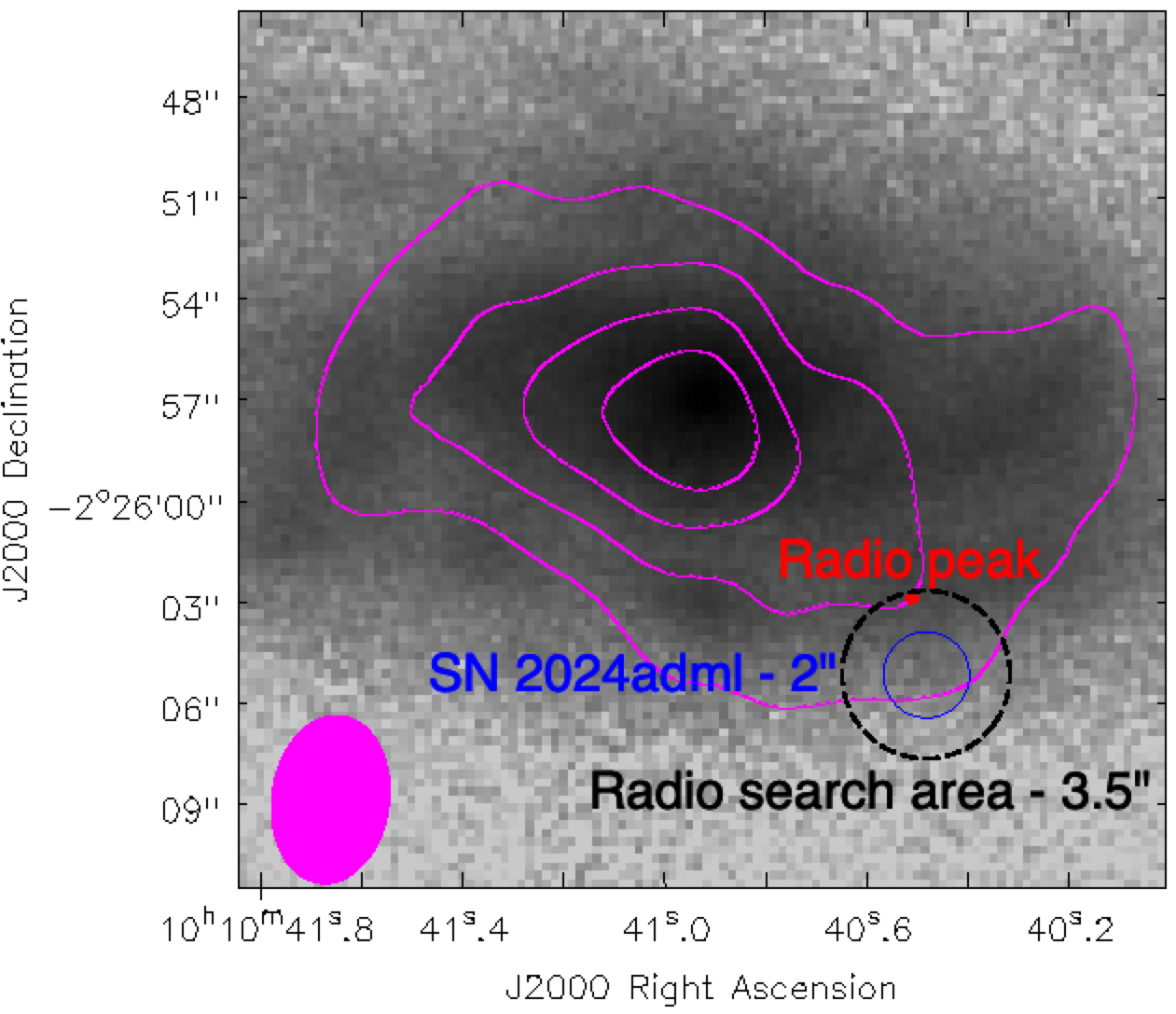}
    \caption{Optical (PanSTARRS-1) image of SN\,2024adml with radio contours mapped in magenta. The radio contours correspond to the third epoch taken $\approx181$ days after the estimated explosion time (at 5.5 \,GHz with the VLA in its C configuration). The black circle is centered at the optical SN position and has a radius equal to the nominal VLA synthesized beam FWHM for that epoch  (3.5\,arcsec). The blue circle has a radius of 2\,arcsec, comparable with the position accuracy of ZTF. The magenta contour lines show that the majority of the radio emission, including the peak (red dot) of the excess radio emission we measure in our VLA search area (black circle) likely originates from the central region of the host galaxy (see \S\ref{sec:host galaxy emission} for further discussion).}
    \label{2024adml}
\end{figure}

\subsection{Radio constraints}
\label{sec:radioanalysis}
As shown in Table~\ref{tab:2}, we obtained confident radio detections ($>3\sigma$ level) for seven SNe in our sample.  Four of these seven (SN\,2022xzc, SN\,2023zeu, SN\,2024abup, SN\,2024adml)  are associated with emission that is likely dominated by host galaxy light  (see Figures \ref{fig:xzc emission}-\ref{2024adml}). The remaining three are radio detections most likely associated with genuine SN radio counterparts. 

As evident from Figure \ref{fig:radio_light}, none of the SNe in our sample for which we exclude radio emission dominated by host galaxy light show evidence for SN\,1998bw-like emission. The three events for which we have a radio counterpart detection (SN\,2020jqm and SN\,2022xxf, both of Type Ic-BL; and SN\,2024rjw, a Type Ic) suggest the presence of either mildly-relativistic ejecta or strong interaction with a dense circumstellar medium (CSM). As we describe in more details in what follows, SN\,2022xxf has a double-peaked optical light curve (Figure \ref{fig:xxf}) which suggests CSM interaction. However, the radio observations do not provide evidence for strong CSM interaction, but rather suggest the presence of mildly-relativistic ejecta. SN\,2020jqm is a double-peaked SN Ic-BL with C-band detections presented in \citet{2023ApJ...953..179C}. SN\,2024rjw appears point-like in our images and shows a late-time rise in flux. While it has been reclassified as a Type Ic SN, we retain it in the sample due to its strong radio emission and light curve behavior, which is indicative of a dense CSM interaction similar to SN\,2020jqm \citep{2023ApJ...953..179C} and PTF11qcj \citep{2019ApJ...872..201P}. 

We also analyzed VLA Sky Survey (VLASS) pre- and post-explosion quick look images available for five of the SNe in our sample (SN\,2020jqm, SN\,2022xzc, SN\,2022crr, SN\,2022xxf, SN\,2023eiw), and pre-explosion images available for four of our SNe (SN\,2023zeu, SN\,2024rjw, SN\,2024abup, SN\,2024adml). The VLASS images revealed no significant ($>3\sigma$) radio detections at the optical SN positions. This is not surprising given that the VLASS rms sensitivity of $\approx0.12$\,mJy at 3\,GHz \citep{2018ApJ...866L..22L,hernández2018searchextendedradiosources} is much shallower than achieved via our deep VLA follow up.

In what follows, we describe the constraint derived via our VLA observations in detail.

\subsubsection{Host-galaxy-dominated radio emission}
\label{sec:host galaxy emission}
SN\,2022xzc displays strong evolution in the observed radio flux across our two epochs. In our first observation, carried out with the VLA in its C configuration, there is evidence for extended emission at 5.5\,GHz (Figure~\ref{fig:xzc emission}), with a $\approx 16 \pm 3.4\,{\rm arcsec}\,\times 6\pm 1.5\,{\rm arcsec}$ emitting region (derived using the \texttt{imfit} task in \texttt{CASA} in a circular region centered on the optical SN position with radius 3.5 arcsec). The measured peak flux density of $\approx 47\,\mu$Jy shows a large discrepancy with the integrated flux of $\approx 409\,\mu$Jy, as expected in the case of extended emission. In our second epoch, carried out with the VLA in its B configuration, the emission appears resolved out and we get a non detection in the searched area. We can estimate a lower limit on the host-galaxy star formation rate (SFR) required to account for the observed radio emission, using the above measured integrated flux density and the following relation \cite{2011ApJ...737...67M}:
\begin{equation}
\label{eq:2}
    \left( \frac{{\rm SFR}_{\rm 1.4\,GHz}}{M_{\odot}{\rm yr}^{-1}} \right) =
    6.35 \times 10^{-29} \left( \frac{L_{\rm 1.4\,GHz}}{\rm{erg\,s}^{-1}{\rm Hz}^{-1}} \right),
\end{equation} 
which implies ${\rm SFR}\gtrsim  1\,M_{\odot}\,{\rm yr}^{-1}$ 
when extrapolating the 1.4\,GHz flux from the one at $\nu_{\rm obs}=5.5$\,GHz using a spectral index $\alpha=-0.7$, i.e.:
\begin{equation}
    L_{1.4\,{\rm GHz}}=4\pi D^{2}_{L} S_{\nu_{\rm obs}} \left(\frac{1.4\,\rm GHz}{\nu_{\rm obs}}\right)^{\alpha}(1+z)^{\alpha-1},
\end{equation} 
where $S_{\nu\rm_{obs}}$ is the integrated radio flux as estimated for the host galaxy. The derived radio SFR value is broadly consistent with the value of $\approx 0.4$\,M$_{\odot}$\,yr$^{-1}$ that we derive from an SED fitting to the NASA/IPAC Extragalactic Database (NED) host galaxy photometry. We caution that due to various degeneracies in the fit, as well systematics related to the inability to capture the full galaxy light in the photometry of nearby objects, all optical SFRs quoted hereafter are only accurate to about a factor of $\approx 3$.  

In the case of SN\,2023zeu, we measure a constant radio flux density over two epochs spanning a time period of more than 500 days post-explosion (see Table~\ref{tab:radioproperties}), and the detected radio emission in the C configuration (Figure~\ref{fig:2023zeu}), is resolved at 5.5\,GHz, with a beam de-convolved size of $7.06\pm 1.1\,{\rm arcsec} \times 0.87\pm0.87\,{\rm arcsec} $ (derived using \texttt{imfit} task in \texttt{CASA} in a circular region centered on the optical SN position with radius 3.5 arcsec). This epoch displays a peak flux density and an integrated flux density of $\approx 36\,\mu$Jy and $\approx 66\mu$Jy, respectively. The estimated radio SFR is $\gtrsim 0.2\,M_{\odot}\,{\rm yr}^{-1}$, broadly compatible with what we obtain from the optical host galaxy light ($\approx 0.1\,M_{\odot}\,{\rm yr}^{-1}$).

For SN\,2024abup, our first two $\approx 5$\,GHz observations carried out (at $\approx22$\,d and $\approx52$\,d since explosion) with the VLA in its most extended  A configuration did not yield a radio detection at 5\,GHz. However, strong radio emission was detected at the same frequency with the VLA in its C configuration. The emission as measured in a circular region of radius 3.5 arcsec centered on the optical SN position is extended (Figure \ref{2024abup}), with an estimated size of $\approx 46\pm0.64$\,arcsec$\times10\pm0.18$\,arcsec. The measured peak flux density is $\approx 232\,\mu$Jy, while the integrated flux is $\approx 6$\,mJy. The last implies a radio SFR rate of $\approx 0.7\,M_{\odot}\,{\rm yr}^{-1}$ (assuming a spectral index as above). Estimating the SFR of the host galaxy of SN\,2024abup from optical light is challenging because its large angular size makes NED photometric estimates unreliable. We obtain an order-of-magnitude estimate of $\approx 0.2\,M_{\odot}\,{\rm yr}^{-1}$ using the NED-reported H-$\alpha$ flux. However, this estimate does not account for dust correction and given the visible dust lane, the optical SFR estimate is likely a factor of a few higher than derived (and hence compatibel with the radio estimate).

Finally, the first two 5.5\,GHz observations of SN\,2024adml with the VLA in its A configuration (at $\approx9$\,d and $\approx34$\,d since explosion) did not show evidence for significant radio emission (see Table~\ref{tab:2}). However, our third observation conducted at $\approx181$\,d post explosion with the VLA in its C configuration shows extended radio emission (Figure \ref{2024adml}) of size $\approx 10\pm0.51\,{\rm arcsec}\times7.7\pm0.35$\,arcsec (derived using \texttt{imfit} task in \texttt{CASA} and a circular region of radius 3.5 arcsec around the optical SN position). The measured peak flux density is $\approx 68\,\mu$Jy, while the integrated flux one is $\approx 450\,\mu$Jy. The last implies a SFR of $\approx 2\,M_{\odot}\,{\rm yr}^{-1}$, to be compared with a value of $\approx 0.4$\,M$_{\odot}\,{\rm yr}^{-1}$ obtained from the optical SED modeling.

\begin{figure*}
        \centering        \includegraphics[width=0.7\textwidth]{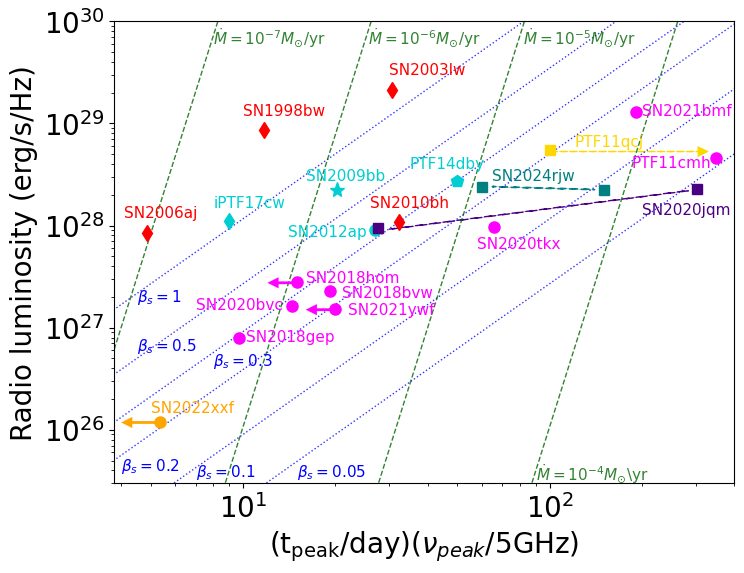}
        \caption{ Properties of the radio-emitting ejecta of the SNe in our sample for which we detect a radio counterpart (orange dot), and those detected in previous studies \citep[magenta dots;][]{2014ApJ...782...42C,Corsi-2016,2017ApJ...847...54C,2023ApJ...953..179C, 2024ApJ...976...71S}, compared with those of GRB-SNe \citep[red diamonds;][]{1998Natur.395..663K,2004Natur.430..648S, 2013ApJ...778...18M, 2006Natur.442.1008C} and of relativistic-to-mildly relativistic SNe Ic-BL discovered independently of a $\gamma$-ray trigger \citep[cyan;][]{2010Natur.463..513S,2017ApJ...847...54C,Milisavljevic_2015}. Radio detections with solid arrows indicate lower limits constraints on the ejecta speeds ($\beta_s$ normalized to c). SN\,2020jqm has a double-peaked radio light curve and we estimate the radio ejecta speed  associated with each of these two peaks. We note that the second radio peak places SN\,2020jqm in the region of the parameter space occupied by radio-loud CSM-interacting SNe similar to PTF11qcj \citep{2014ApJ...782...42C}. SN\,2024rjw is a SN Ic with radio emission that supports the interpretation of this event as a CSM-interacting one based on the double-peaked optical light curve. See \S\ref{sec:radio_properties} for further discussion.}
        \label{fig:radio_speed}
\end{figure*}

\subsubsection{Properties of the radio-emitting ejecta}
\label{sec:radio_properties}
Here, we constrain the physical properties of the radio-emitting ejecta of the SNe for which we have radio detections (not associated with host galaxy light), within the synchrotron self-absorption (SSA) model for radio SNe \citep{1998ApJ...499..810C}. In this model, the measured radio peak frequency and flux provide estimates of the emitting region's size (and hence velocity), as well as the progenitor's mass-loss rate. Starting from Equations (11) and (13) of \citet{1998ApJ...499..810C}:
\begin{equation}
\label{eq 3}
\begin{aligned}
    R_p \approx 8.8 \times 10^{15} \, \mathrm{cm} \left( \frac{\eta}{2 \alpha} \right)^{1 / (2p + 13)}
    \left( \frac{F_p}{\mathrm{\mu Jy}} \right)^{(p + 6) / (2p + 13)} \\
    \times \left( \frac{d_L}{\mathrm{Mpc}} \right)^{(2p + 12) / (2p + 13)}
    \left( \frac{\nu_p}{5 \, \mathrm{GHz}} \right)^{-1},
\end{aligned}
\end{equation}
where $\alpha \approx 1$ is the ratio of relativistic electron energy density to magnetic energy density, $F_p$ is the flux density at the SSA peak, $\nu_p$ is the SSA frequency, and $R/\eta$ is the thickness of the radiating electron shell, with $\eta \equiv f^{-1}$ the inverse of the filling factor $f$ of the synchrotron-emitting region \citep{1998ApJ...499..810C,2005ApJ...621..908S}.

By setting $R_p \approx \nu_s t_p$ in Equation~\ref{eq 3}, and using $L_p \approx 4 \pi d_L^2 F_p$, we derive:
\begin{equation}
\label{eq 4}
\begin{aligned}
    \left( \frac{L_p}{\mathrm{erg} \, \mathrm{s}^{-1} \, \mathrm{Hz}^{-1}} \right) 
    \approx 1.2 \times 10^{27} \left( \frac{\beta_s}{3.4} \right)^{(2p + 13) / (p + 6)} \\
    \times \left( \frac{\eta}{2 \alpha} \right)^{-1 / (p + 6)}
    \left( \frac{\nu_p}{5 \, \mathrm{GHz}} \frac{t_p}{1 \, \mathrm{day}} \right)^{(2p + 13) / (p + 6)},
\end{aligned}
\end{equation}
where $\beta_s = \nu_s/c$. In Figure~\ref{fig:radio_speed}, we plot this relation for various values of $\beta_s$, assuming $p = 3$, $\eta = 2$, and $\alpha = 1$. We note that $\eta$ is typically assumed to be in the range 2--10 \citep{1998ApJ...499..810C,2005ApJ...621..908S}. Adopting  different values of $\eta$ in this range would change our estimates by $\lesssim 10\%$. Relativistic events like SN\,1998bw lie above the $\beta_s \gtrsim 1$ regime, outside the validity of the non-relativistic assumptions in these equations.

To estimate the speed of the radio-emitting ejecta for the three SNe in our sample with radio detections, we apply Equation~\ref{eq 3} using our observed values and plot the results in Figure~\ref{fig:radio_speed}. For sources where the peak flux occurred at the first epoch of observation, we mark them with arrows to indicate that the true peak could lie at earlier times (further to the left on the plot).

SN\,2020jqm (Ic-BL) and SN\,2024rjw (Ic)  both show ejecta speeds $\lesssim 0.5$\,c and point to late-time peaking radio emission compatible with strong circumstellar interaction (Figure~\ref{fig:radio_speed}), similarly to what observed in other SNe Ic-BL \citep[e.g.,][]{2014ApJ...782...42C} and Ib/c events \citep[e.g.,][]{2012ApJ...752...17W}. As we show below, this interpretation is reinforced by the estimated progenitor mass-loss rates.

We can also estimate the progenitor’s mass-loss rate using SSA theory. From Equations (12) and (14) of \citet{1998ApJ...499..810C}, the magnetic field is:
\begin{equation}
\begin{aligned}
    B_p \approx 0.58 \, \mathrm{G} \left( \frac{\eta}{2 \alpha} \right)^{4 / (2p + 13)}    \left( \frac{F_p}{\mathrm{Jy}} \right)^{-2 / (2p + 13)} \\
    \times \left( \frac{d_L}{\mathrm{Mpc}} \right)^{-4 / (2p + 13)}
    \left( \frac{\nu_p}{5 \, \mathrm{GHz}} \right),
\end{aligned}
\end{equation}

Assuming the shock expands into a circumstellar medium (CSM) with density:
\begin{equation}
    \rho \approx 5 \times 10^{11} \, \mathrm{g} \, \mathrm{cm}^{-3} \, A_* \, R^{-2},
\end{equation}

Also assuming a fraction $\epsilon_B$ of the energy density $\rho v_s^2$ goes into magnetic fields, we obtain:
\begin{equation}
\label{eq 8}
    \frac{B_p^2}{8 \pi} = \epsilon_B \rho \nu_s^2 = \epsilon_B \rho R_p^2 t_p^{-2}.
\end{equation}

Solving for $L_p$ yields:
\begin{equation}
\label{eq 9}
\begin{aligned}
    \left( \frac{L_p}{\mathrm{erg} \, \mathrm{s}^{-1} \, \mathrm{Hz}^{-1}} \right)
    \approx 1.2 \times 10^{27} \left( \frac{\eta}{2 \alpha} \right)^2
    \left( \frac{\nu_p}{5 \, \mathrm{GHz}} \frac{t_p}{1 \, \mathrm{day}} \right)^{\frac{(2p + 13)}{2} }\\
    \times \left( 5 \times 10^3 \epsilon_B A_* \right)^{-(2p + 13)/4}.
\end{aligned}
\end{equation}

In Figure~\ref{fig:radio_speed}, we overlay this relation with green dashed lines for various $\dot{M}$ values, assuming $p = 3$, $\eta = 2$, $\alpha = 1$, $\epsilon_B = 0.33$, and $v_w = 1000\,\mathrm{km\,s^{-1}}$. We find that relativistic events like SN\,1998bw favor lower inferred mass-loss rates, while CSM interacting events such as PTF\,11qcj, SN\,202jqm, and SN\,2024rjw favor larger mass-loss rates. While Equation~\ref{eq 8} depends on assumed values for $\eta$, $\epsilon_B$, and $v_w$, the trend in $\dot{M}$ holds across typical parameter choices.

\begin{table*}
\begin{center}
\caption{Properties of the radio-emitting ejecta for the SNe in our sample with detected radio counterparts. From left to right, we report the SN name, the estimated ejecta speed normalized to $c$ ($\beta_s$), the progenitor mass-loss rate ($\dot{M}$), the energy coupled to the fastest (radio-emitting) ejecta ($E_r$), and the ratio between $E_r$ and the total explosion kinetic energy $E_k$ (from optical modeling; see \S\ref{sec:radioanalysis}). For SN\,2024rjw, we list results for the two brightest detections at $t=53$ and $t=133$ days. See Table~\ref{tab:2} for details.
\label{tab:radioproperties}}
\begin{tabular}{lcccc}
\toprule
\toprule
SN & $\beta_s$ & $\dot{M}$ ($M_\odot$\,yr$^{-1}$) & $E_r$ (erg) & $E_r/E_k$ \\
\midrule
2022xxf & $\gtrsim 0.2$ & $\lesssim 3\times10^{-7}$ & $\lesssim 10^{46}$ & $<0.001\%$ \\
2024rjw & $\gtrsim 0.2$ ($ 0.09$) & $\lesssim 10^{-5}$ ($ 7\times10^{-5}$) & $\lesssim 6\times10^{48}$ ($ 5\times10^{48}$) & $< 0.2\%$ \\
\bottomrule
\end{tabular}
\end{center}
\end{table*}

Finally, we estimate the energy coupled to the fastest ejecta using \citet{2006ApJ...651.1005S}:
\begin{equation}
    E_r \approx \frac{4 \pi R_p^3}{\eta} \frac{B_p^2}{8 \pi \epsilon_B} = \frac{R_p^3}{\eta} \frac{B_p^2}{2 \epsilon_B}.
\end{equation}
In Table~\ref{tab:radioproperties}, we summarize the derived properties for the two detected radio SNe (SN\,2022xxf, SN\,2024rjw), SN\,2020jqm was covered in \citep{2023ApJ...953..179C}. All show evidence for energies in the radio-emitting ejecta (and progenitor mass-loss rates) that are significantly lower (higher) than those of GRB-SNe such as: SN\,1998bw: $\dot{M} \approx 2.5 \times 10^{-7} M_\odot\,\text{yr}^{-1}$, $E_r \approx (1 - 10) \times 10^{49}$ erg \citep{1999ApJ...526..716L}; SN\,2009bb: $\dot{M} \approx 2 \times 10^{-6} M_\odot\,\text{yr}^{-1}$, $E_r \approx 1.3 \times 10^{49}$ erg \citep{2010Natur.463..513S}; GRB\,100316D/SN\,2010bh: $\dot{M} \approx (0.4 - 1) \times 10^{-5} M_\odot\,\text{yr}^{-1}$, $E_r \approx (0.3 - 4) \times 10^{49}$ erg \citep{2013ApJ...778...18M}.

\begin{figure*}
    \centering
    \hbox{
  \includegraphics[width=8.5cm]{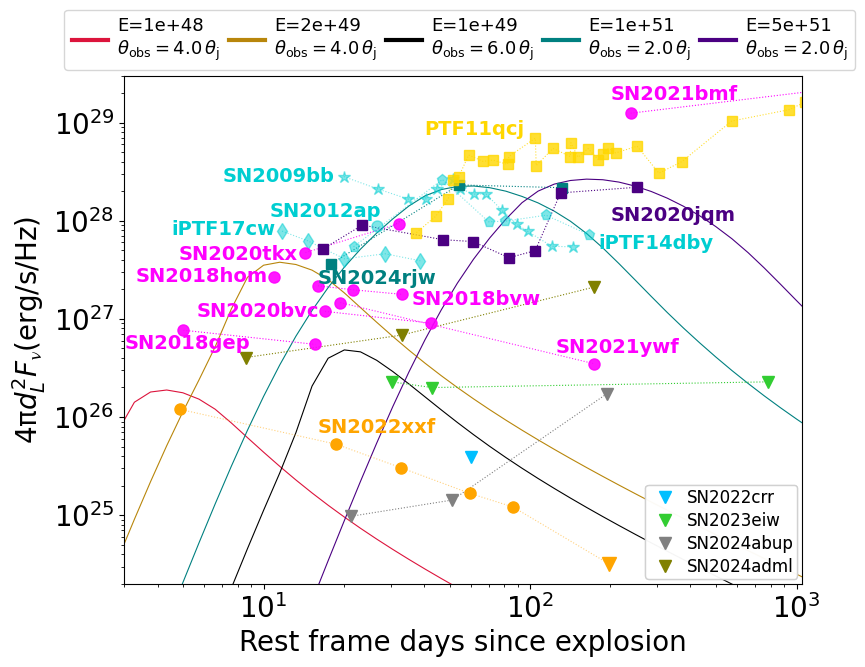} 
    \includegraphics[width=9.5cm]{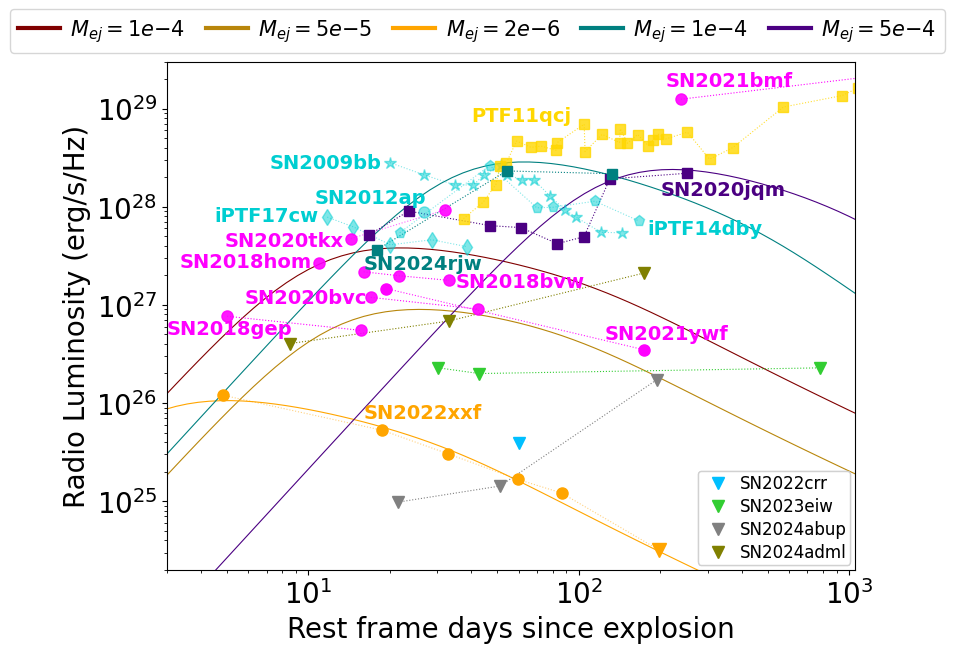}
  }
    \caption{ LEFT: Off-axis GRB afterglow model light curves using \texttt{afterglowpy} \citep{2020ApJ...896..166R}, similar to Figure \ref{fig:X-ray LCs comparison} but with lower values of the energy $E$ and larger off-axis angles (see legend). Overall, these models show post-peak decay that is faster than what typically observed in the radio. RIGHT:  \texttt{afterglowpy} cocoon models (solid lines) with input parameters $\beta_s=0.71-0.79$, $n=1$\,cm$^{-3}$(uniform medium), and ejecta masses $M_{ej}= 10^{-5}M_{\odot}- 2\times10^{-6}M_{\odot}$. For SN2022xxf, a cocoon model with the following parameters fits the observed radio data remarkably well: $\beta_s=0.79$, $M_{ej}= 2\times10^{-7}M_{\odot}$, see \S\ref{sec:cocoon potential} for further discussion. We also show cocoon models that are compatible with SN\,2024rjw and may explain the late-time re-brightening of SN\,2020jqm. However, we caution that an interpretation of these two SNe within the cocoon model requires extended late-time radio follow up.}
        \label{fig:radio cocoons}
\end{figure*}

\subsubsection{The rate of SN\,1998bw-like SNe Ic-BL}
\label{sec:1998bw invertigation}

SNe Ic-BL constitute about $5\%$ of all core-collapse SNe \citep{2011MNRAS.412.1441L, 2017PASP..129e4201S}.  Based on the ZTF sample, we are starting to constrain the rates of the brightest Ic-BL events with greater accuracy. Specifically, we derive volumetric rates of $\approx 80 - 200\,\text{Gpc}^{-3}\,\text{yr}^{-1}$ for events with peak $r$-band absolute magnitudes in the range $-19\,{\rm mag} \le M_r < -18.5$\,mag, and  $\approx 25 - 90\,\text{Gpc}^{-3}\,\text{yr}^{-1}$ for events with $M_r < -19$\,mag). Because the median peak absolute magnitude of the SNe in our sample is $M_r = -17.82$, and only one SN in our sample falls in the $M_r < -18.5$ range, hereafter we adopt a cumulative volumetric rate of SNe Ic-BL of $\approx 740 - 2600\,\text{Gpc}^{-3}\,\text{yr}^{-1}$, which is based on the whole ZTF sample, D. Perley et al. 2026 (in preparation).

The most stringent constraints on the rate of 1998bw-like events among SNe Ic-BL have been set by \citet{2023ApJ...953..179C}, who have shown that 1998bw-like events (defined here as events with radio emission
observationally similar to SN 1998bw) constitute less than $19\%$ (at 99.865\% confidence) of the SN Ic-BL population. With the additional observations presented here (7 new SNe Ic-BL for which we can exclude radio emission observationally similar to SN\,1998bw/GRB\,980425, see Figure \ref{fig:radio_light}), our sample of radio-monitored SNe Ic-BL has expanded to 41 in total. This lowers the 99.865\% confidence upper limit on the fraction of 1998bw-like events to $<6.61/41 \approx 16\%$, or a volumetric rate of $R_{\rm 1998bw-like} < 118 - 416\,\text{Gpc}^{-3}\,\text{yr}^{-1}$. Note that we have used the fact that the Poisson 99.865\% confidence (or $3\sigma$ Gaussian
equivalent for a single-sided distribution) upper limit on zero
SNe compatible with SN 1998bw is $\approx 6.61$.

If we assume that all or the majority of low-luminosity GRBs have radio luminosities comparable to that of SN\,1998bw, we can use the observed rate of low-luminosity GRBs ($R_{\rm LLGRB}$) and our derived constraints on the rate of 1998bw-like events ($R_{\rm 1998bw-like}$) to infer the low-luminosity GRB beaming angle:
\begin{equation}
R_{\rm LLGRB} = (1-\cos\theta)R_{\rm 1998bw-like}.
\end{equation}
Hereafter, we adopt $R_{\rm LLGRB} \approx 164-325\,\text{Gpc}^{-3}\,\text{yr}^{-1}$ based on \citet{Sun_2015} and \citet{Liang_2007}. Our upper limit of $R_{\rm 1998bw-like} < 118 - 416\,\text{Gpc}^{-3}\,\text{yr}^{-1}$ sets a lower limit on the beaming angle $\theta$ :
\begin{equation}
\cos\theta = 1-\frac{R_{\rm LLGRB}}{R_{\rm 1998bw-like}} \lesssim  1-\frac{164\,\text{Gpc}^{-3}\,\text{yr}^{-1}}{416\,\text{Gpc}^{-3}\,\text{yr}^{-1}},
\end{equation}
\begin{equation}
\theta  \gtrsim 52\,{\rm deg}.
\end{equation}
It is clear that our radio observations increasingly disfavor the hypothesis that all low-luminosity GRBs are associated with 1998bw-like fast ejecta. 
On the other hand, radio emission similar to SN\,2006aj, associated with GRB\,060218, cannot be excluded for most of the SNe Ic-BL in our sample (see Figure \ref{fig:radio_light}), and would require faster radio follow ups \citep{2023ApJ...953..179C}. Indeed, the working assumption in our analysis is that low-luminosity GRBs constitute a class of relativistic explosions with relatively small beaming factors \citep{Liang_2007}, and that SN\,1998bw is observationally representative of the bulk properties of this class. Under these assumptions, their radio light curves are expected to be less affected by viewing-angle effects. However, if the majority of the low-luminosity GRB population exhibited radio light curves more akin to SN\,2006aj i.e., significantly fainter and more rapidly evolving, our study would not be able to probe them.

We emphasize that a systematic approach to radio follow-up campaigns would be highly beneficial, providing a unique opportunity to tighten existing constraints to the point where SN 1998bw-like emission could be firmly excluded in all SNe Ic-BL. To illustrate this potential, we examined the ZTF Bright Transient Survey catalog \citep{2020ApJ...895...32F, 2020ApJ...904...35P}, which contains approximately 90 SNe Ic-BL with redshifts $z\lesssim0.2$, detected between 2018 - 2025. If sufficient VLA observing time had been available to monitor each of these events with the same sensitivity as achieved in this work (i.e., $\approx 50\,\mu\text{Jy}$ at $5\sigma$ for 6\,GHz observations of sources at $z\lesssim0.2$), the resulting upper limit on the fraction of 1998bw-like SNe Ic-BL could have been as low as $<6.61/90\approx7\%$, corresponding to an event rate of $R_{\rm 1998bw-like}\lesssim 180\,{\rm Gpc}^{-3}\,{\rm yr}^{-1}$. This value is comparable to the lowest estimates of the low-luminosity GRB rate \citep{Sun_2015}, underscoring the importance of systematic, high-sensitivity radio follow-up for constraining the diversity of engine-driven explosions.

\subsection{Off-axis jets or cocoons?}
\label{sec:cocoon potential}
As evident from Figure~\ref{fig:radio_light}, while our radio follow-up campaigns seem to be consistently excluding SN\,1998bw-like radio emission for the majority of SNe Ic-BL, several events are as radio loud as SN\,1998bw but their emission peaks much later. SN\,2024rjw, classified as a Type Ic SN and presented here for the first time, shows persistent radio brightness beyond 100 days post-explosion (see Figure~\ref{fig:radio_light}). SN\,2020jqm, analyzed in both \citet{2023ApJ...953..179C} and \citet{2024ApJ...976...71S}, exhibits a double-peaked radio light curve, with the second, stronger peak appearing more than 200 days after the estimated explosion time. A similar source, SN\,2021bmf, discussed in \citet{2024ApJ...976...71S} and \citet{Anand_2024}, displays late-time radio emission emerging after $\sim$200 days and peaking after the 1000\,d since explosion. Overall, is clear from Figure \ref{fig:radio_speed} that these radio-loud and late-time peaking events show low ejecta speeds and higher values of the progenitor mass-loss rate, suggesting that strong CSM interaction contributes substantially to their radio emission. 

Also evident from Figure \ref{fig:radio_speed} is the fact that, in addition to the late-time-peaking radio-loud events mentioned above, a second class of radio-emitting SNe Ic-BL is starting to emerge. Namely, SNe Ic-BL with radio luminosities at least one order of magnitude dimmer than SN\,1998bw, and with smoothly decaying light curves, (SN2018gep, SN2018bvw, SN2020bvc, SN2021ywf, and SN2022xxf; see Figure \ref{fig:radio_light}). These events suggest fastest ejecta speeds in between $20\%c$ and $50\%c$, and mass-loss rates $\lesssim 10^{-5}\,M_{\odot}$\,yr$^{-1}$. 

In what follows, we ask whether these two classes of radio-emitting SNe Ic-BL that do not resemble the GRB-associated SN\,1998bw, can be interpreted within either off-axis and/or lower-energy top-hat jet models (i.e., jet models not already excluded via our X-ray observations), or models characterized by quasi-spherical ejecta with velocity stratification, accelerated through interaction with possibly choked jets.

In Figure~\ref{fig:radio cocoons} we compare our data with the two families of models above. While largely off-axis or lower energy top-hat jets (right panel) may contribute to the late-time emission of some of the radio-loud and late-time peaking SNe (left panel), it is clear that these models predict a post-peak radio light curve decay generally faster than what observed in our data. On the other hand, cocoon  models with ejecta masses in the range $5\times10^{-6}\,M_\odot$ to $10^{-5}\,M_\odot$ are broadly compatible with some of the radio-emitting SNe in our sample. More specifically, cocoon models with with ejecta masses $1-5\times10^{-5}\,M_\odot$, constant density medium $n = 0.005-0.015\,\mathrm{cm}^{-3}$, microphysical parameters $\epsilon_e=0.33$ and $\epsilon_B=0.33$, and maximum ejecta speed of $\approx 0.74-0.8c$, may explain or contribute to the late-time emission observed in SN\,2024rjw and SN\,2020jqm. Long-term radio monitoring is critical to distinguish a potential top-hat off-axis jet contribution to these events from a cocoon one, given that cocoon emission has much slower rate of temporal decay at very late times, for more information for late time Ic-BL SNe behavior we refer the reader to \cite{schroeder2025latetimeradiosearchhighly}. 

\begin{figure}
    \begin{center}
    \includegraphics[width=0.47\textwidth]{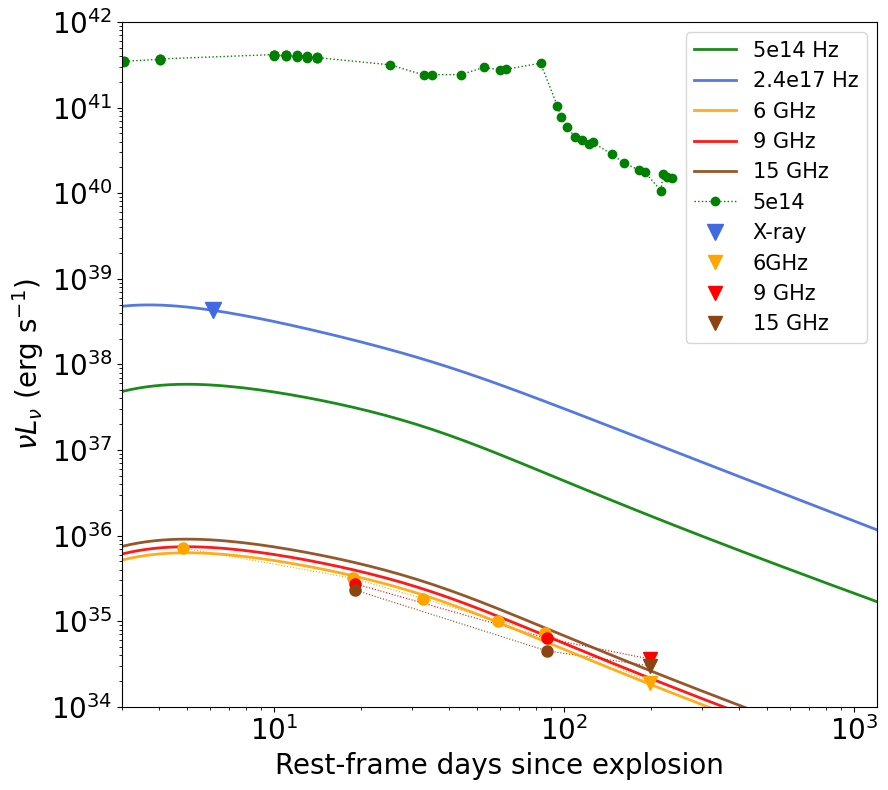}
    \caption{SN\,2022xxf observations are compared with predictions from a cocoon model with ejecta mass $2\times10^{-7}\,M_\odot$, ISM-like density $n = 1\,\mathrm{cm}^{-3}$(uniform medium), microphysical parameters $\epsilon_e=0.1$ and $\epsilon_B=0.01$, and maximum ejecta velocity of $ \approx 0.8c$ and minimum ejecta velocity of $ \approx 0.1c$. This model matches the radio data we collected for SN\,2022xxf data reasonably well (gold). It also agrees with the upper limit we collected in the X-rays (blue). The contribution of the cocoon model (green solid line) to the optical light curve of SN\,2022xxf (green dots and dotted line) is negligible, and hence not in contrast with the idea that the double-peaked optical light curve of SN\,2022xxf is powered by a combination of regular $^{56}$Ni decay (powering the first optical hump) and ejecta-CSM interaction \citep[powering the second optical peak;][]{2023A&A...678A.209K}. See \S\ref{sec:cocoon potential} for further discussion.}
    \label{fig:XXF bands comparison }
    \end{center}
\end{figure}

Interestingly, as highlighted in Figure~\ref{fig:XXF bands comparison }, a cocoon model with ejecta mass $2\times10^{-7}\,M_\odot$, ISM-like density $n = 1\,\mathrm{cm}^{-3}$, microphysical parameters $\epsilon_e=0.1$ and $\epsilon_B=0.01$, and maximum ejecta velocity of $\approx 0.8c$ matches the radio data we collected for SN\,2022xxf data rather well. This model also agrees with the upper limit we collected in the X-rays. The cocoon contribution to the optical light curve of SN\,2022xxf is negligible, and hence not in contrast with the idea that the double-peaked optical light curve of SN\,2022xxf is powered by a combination of regular $^{56}$Ni decay (powering the first optical hump) and ejecta-CSM interaction \citep[powering the second optical peak;][]{2023A&A...678A.209K}. In this scenario, the radio emission we observe suggests the presence of a mildly-relativistic ejecta tail or cocoon, formed in the interaction between the SN shock and the CSM. Late-time radio monitoring ($\approx$1,000 days post explosion) could constrain the presence of potential radio re-brightenings from interaction with higher-density CSM shells and help distinguish a simple cocoon component powering the radio emission from a more complex CSM interaction model with shells of different densities  \citep{2019ApJ...872..201P}. 

\section{Multi-messenger detection prospects}
\label{sec:multimessenger}

\begin{figure}[t]
\includegraphics[width=0.47\textwidth]{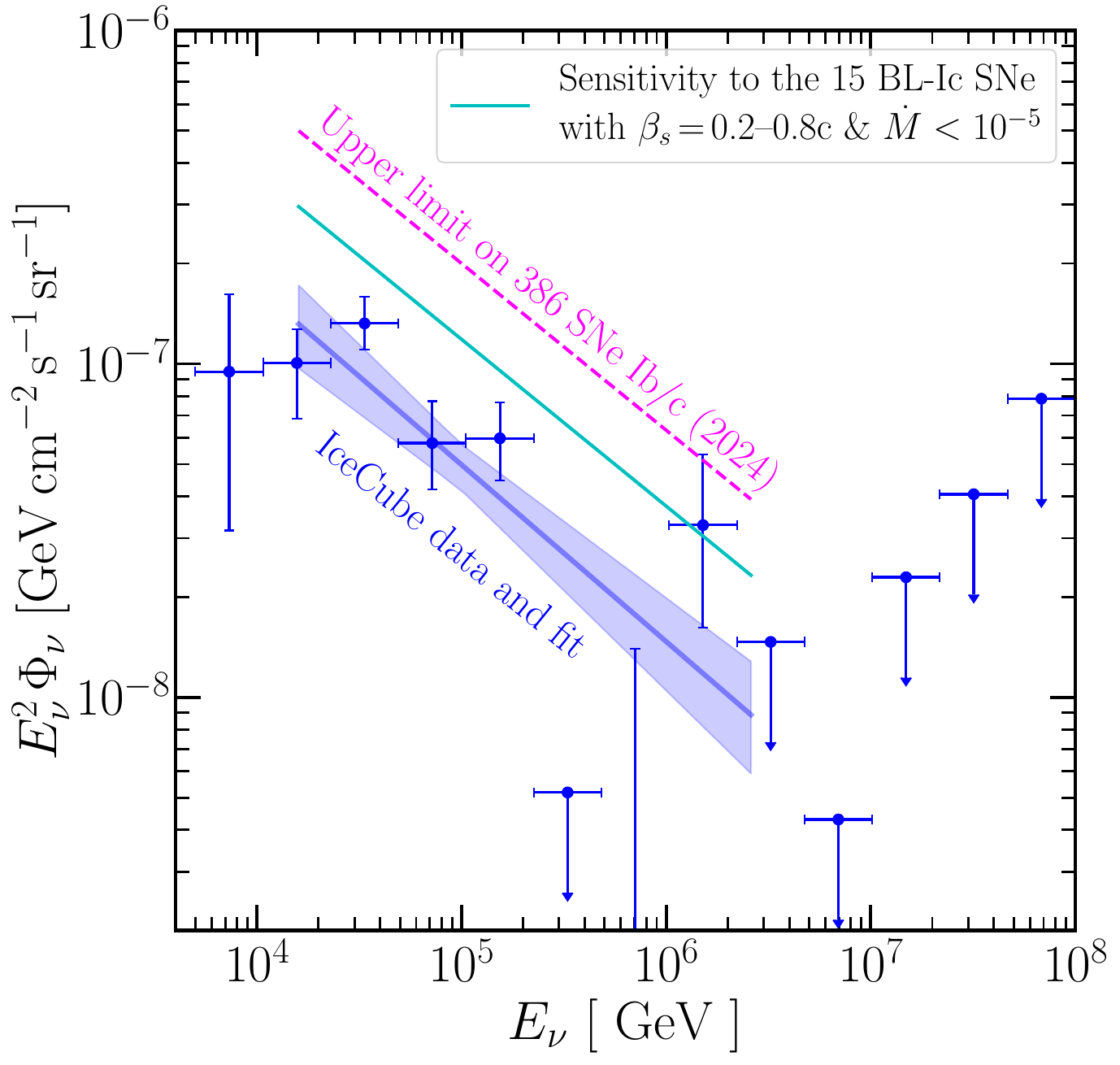}
\caption{
Our estimated HE neutrino sensitivity of an IceCube-like neutrino telescope to the 15 SNe Ic-BL with radio ejecta speeds, $\beta_s$, between 0.2--0.8c and mass loss rate, $\dot{M}$, below $10^{-5}$ in Figure~\ref{fig:radio_speed}, along with the upper limits from previous analysis of 386 SNe Ib/c from 2008–2018~\citep{2024PhRvD.109j3041C} at 95\% confidence level, compared with the diffuse neutrino flux measured by IceCube~\citep{IceCube:2020acn}. 
For illustration purposes, the sensitivity and upper limit lines assume $f_{\rm jet} = 1$ (for illustrative purposes), meaning all SNe are assumed to host choked jets aligned toward Earth.
See text for details.
}
\label{fig_HEnu}
\end{figure}

SNe associated with cocoons from choked jets have been proposed as promising sources of high-energy (HE) neutrinos. While successful jets power classical long GRBs, mildly relativistic or failed jets can remain hidden from gamma-ray observatories due to optical thickness or insufficient breakout, but still accelerate hadrons that pproduce neutrinos via photohadronic interactions. These neutrinos, being weakly interacting, can escape dense stellar envelopes and serve as unique messengers of jet activity inside exploding massive stars~\citep{2016PhRvD..93h3003S}.

Several searches using IceCube data have investigated possible neutrino emission from core-collapse SNe, especially those of Type Ib/c~\citep{2016PhRvD..93h3003S, 2018JCAP...01..025S, 2018JCAP...12..008E, 2023ApJ...949L..12A, 2024PhRvD.109j3041C}. Due to the limited number of detected SNe Ic-BL at the time, none of these studies present dedicated analyses focusing solely on SNe Ic-BL.

We estimate the sensitivity of IceCube to the 
15 SNe Ic-BL with radio ejecta speeds, $\beta_s$, between 0.2--0.8c and mass loss rate, $\dot{M}$, below $10^{-5}$ in Figure~\ref{fig:radio_speed}, 
from a stacking likelihood analysis based on \citet{2024PhRvD.109j3041C} and \citet{Zhou:2021rhl} of these SNe and mock IceCube data.
The mock IceCube data is based on IceCube public data~\citep{IceCube:2021xar}, which spans from 2008 to 2018, a different time period from our SN sample, allowing us to treat them as background-only (time-independent) in the vicinity of each SN's sky position.

Figure~\ref{fig_HEnu} shows our estimated HE neutrino sensitivity (magenta line) of IceCube to the 
15 SNe Ic-BL in this work, along with upper limits from a previous analysis of 386 SNe Ib/c (cyan line)~\citep{2024PhRvD.109j3041C} at 95\% confidence level, compared with IceCube's diffuse astrophysical neutrino flux measurement~\citep[blue;][]{IceCube:2020acn}. We note that a recent measurement obtained a very similar result~\citep{IceCube:2024fxo}. 
Importantly, the sensitivity based on the 
15 SNe Ic-BL is more than twice as strong as the upper limit derived from the 386 SNe Ib/c sample~\citep{2024PhRvD.109j3041C}, and approaches IceCube’s measurement~\citep{IceCube:2020acn}. 
This is because the 
15 SNe Ic-BL are, on average, significantly closer to us, and their small number reduces the cumulative foreground, resulting in a higher signal-to-noise ratio compared to the 386 SNe Ib/c. Our results indicate that these 15 SNe Ic-BL offer excellent discovery potential if targeted in a dedicated search for HE neutrino emission using IceCube data. We note that the sensitivity and upper limit lines in Figure~\ref{fig_HEnu} assume $f_{\rm jet} = 1$, meaning all SNe are assumed to host choked jets aligned toward Earth. In practice, for a large sample of Ic-BL selected regardless of the radio emission properties and representative of the Ic-BL class as a whole, we expect $f_{\rm jet} < 10\%$, which would scale the sensitivity curves upward by a factor of $\gtrsim 10$. 
This highlights the importance of a large sample as well as radio monitoring for identifying events that are more likely to be associated with cocoons based on their radio properties.

\section{Summary and Outlook}\label{sec:conclusion}

We presented radio and X-ray follow-up observations of eight SNe Ic-BL and one SN Ic (SN\,2024rjw) that are part of the ZTF sample. Our results confirm that 1998bw-like SNe are intrinsically rare, and favor the idea that SNe Ic-BL constitute a diverse population of stellar explosions, powered by a range of central engines and circumstellar environments. 
The continued discovery of mildly relativistic SNe with radio luminosities between $\sim10^{25}\,\mathrm{erg\,s^{-1}\,Hz^{-1}}$ and $\sim3\times10^{28}\,\mathrm{erg\,s^{-1}\,Hz^{-1}}$ supports the idea that radio-emitting stripped-envelope SNe—consistent with off-axis jets or cocoon emission from choked jets—may be more common.
We demonstrated that at least one out of our sample of nine SNe is fully compatible a cocoon light curve model, we also note that, across our campaigns, we have identified only one event (SN\,2022xxf) that falls into the parameter space of SN\,2006aj-like explosions, characterized by faint, early-time-peaking radio emission just days after explosion. This detection motivates the possibility that more SNe Ic-BL could harbor GRB\,060218/SN\,2006aj-like emission, but are missed due to the scarcity of early-time radio follow-up. As also highlighted by \citet{2023ApJ...953..179C}, rapid spectroscopy and deep radio observations within $\lesssim5$ days of explosion are essential to capture these elusive events.

In the future, the Legacy Survey of Space and Time (LSST) conducted by the Vera C. Rubin Observatory \citep{2019ApJ...873..111I} is poised to revolutionize transient discovery. The upcoming Data Preview 2 (DP2) in May 2026 will provide a reprocessing of all commissioning data, and once fully operational, Rubin is expected to discover $\sim1{,}000{,}000$ supernovae per year, \citep{graham2024lsstalerts}. With a sample of this size we expect systematic VLA follow up to provide key evidence for a SN\,1998bw-like candidate or call into question the currently accepted event rate of low-luminosity GRBs (\S\ref{sec:1998bw invertigation}).  Fast and accurate spectroscopic classification will be critical to fully exploit this discovery potential by triggering follow up in radio and X-rays. Rubin's data, combined with coordinated radio campaigns, will offer unparalleled opportunities to investigate jet formation, CSM structure, and the diverse end of life scenarios for massive stars.

\nocite{VanderWalt2019}
\nocite{Coughlin_2023}
\nocite{murphy2025dawesreview13new}

\appendix
\label{sec:appendix}

\section{SN\,2020\lowercase{jqm} (ZTF20\lowercase{aazkjfv})}
    We refer the reader to \citet{2023ApJ...953..179C}/\citet{2024ApJ...976...71S} for details about this SN Ic-BL. Its P48 light curve and the spectrum used for classification are shown in Figures \ref{fig:1} and \ref{fig:2}, respectively. 

\section{SN\,2021\lowercase{ywf} (ZTF21\lowercase{acbnfos})}
    We refer the reader to \citet{2023ApJ...953..179C}, \citet{Anand_2024}, and \citet{2024ApJ...976...71S} for details about this SN Ic-BL. Its P48 light curve and the spectrum used for classification are shown in Figures \ref{fig:1} and \ref{fig:2}, respectively.

\section{SN\,2022\lowercase{xzc} (ZTF22\lowercase{abnpsou})}
    Our first ZTF photometry of SN 2022xzc was obtained on 2022 October 17 (MJD 59869.53) with P48. The first detection was in \textit{r} band, with host-subtracted magnitude of $19.17\pm0.16$mag, at $\alpha$  = $12^\text{h}01^\text{m}12^\text{s}.58$, $\delta$ = $+\ang{22}36'55{''}.3$ (J2000). The Object was first reported to the TNS by \cite{2022xzc} on October 18, and was first detected by \cite{2022xzc} on October 17 at \textit{r}=18.17 mag \citep{2022xzc_detection}. The last ZTF non-detection was on 2022 October 17 at \textit{r} $>$ 18.4 mag. The transient was classified as an SN Type Ic-BL by \citep{2022xzc_class} with spectra taken with the Alhambra Faint Object Spectrograph and Camera (ALFOSC) on the Nordic Optical Telescope (NOT) shown in Figure \ref{fig:2}. It has  a measured redshift of \textit{z} = 0.027 and was found on the edge of its host galaxy. 

\section{SN\,2022\lowercase{crr} (ZTF\lowercase{22aabgazg})}
    Our first ZTF photometry of SN 2022crr was obtained on 2022 February 18 (MJD 59628.50) with P48. The first detection was in \textit{g} band, with host-subtracted magnitude of $17.76\pm0.09$mag, at $\alpha$  = $15^\text{h}24^\text{m}49^\text{s}.132$, $\delta$ = $-\ang{21}23'21{''}.73$ (J2000). The Object was first reported to the TNS by \cite{2022crr} on February 18, and was first detected by \cite{2022crr} on February 18 at \textit{orange-ATLAS}=17.97 mag \citet{2022crr_detec}. The last ZTF non-detection was on 2022 April 8 at \textit{g} $>$ 19.69 mag, and the last ATLAS nondetection was on 2022 February 17 at \textit{o} $>$ 18.55 ABmag. The transient was classified as an SN Type Ic-BL by \citet{2022crr_class} with spectra taken from the Low-Resolution Imaging Spectrograph (LRIS) on the Keck I Telescope in Hawaii, shown in figure \ref{fig:2}, and found near host galaxy with measured redshift of \textit{z} = 0.0188. We also note this source was used in two recent papers, Keck Infrared Spectra data release 1 \cite{2024PASP..136a4201T}, and Velocity evolution of broad lined Ic in \citet{2024arXiv241111503F}.

\section{SN\,2022\lowercase{xxf} (ZTF22\lowercase{abnvurz})}
    SN 2022xxf is a well studied transient with double peaked optical light curves, which can be seen in Figure \ref{fig:1}. For detailed information about the spectra and progenitor models of this source we refer the reader to \citet{2023A&A...678A.209K}. 
    We used spectra taken from the Double Spectrograph (DBSP) on the Palomar 200-inch Hale Telescope (P200) shown in Figure \ref{fig:2}.

\section{SN\,2023\lowercase{eiw} (ZTF19\lowercase{aawhzsh})}
    Our first ZTF photometry of SN 2023eiw  was obtained on 2023 March 29 (MJD 60032.15) with P48. The first detection was in \textit{g} band, with host-subtracted magnitude of $18.92\pm0.10$mag, at $\alpha$  = $12^\text{h}28^\text{m}46^\text{s}.200$, $\delta$ = $+\ang{46}31'15{''}.64$ (J2000). The Object was first reported to the TNS by \cite{2023eiw} on March 31, and was first detected by \cite{2023eiw} on March 30 at \textit{i}=18.53 mag \citet{2023eiw_detec}. The last ZTF non-detection was on 2023 March 26 at \textit{g} $>$ 20.66 mag. The transient was classified as an SN Type Ic-BL by \citet{2023eiw_class} with spectra taken by the Spectral Energy Distribution Machine (SEDM) attached to Palomar 60-inch telescope (P60). Shown in figure \ref{fig:2} is spectra taken form the Alhambra Faint Object Spectrograph and Camera (ALFOSC) on the Nordic Optical Telescope (NOT). The supernova host-galaxy has a known redshift of \textit{z} = 0.025, measured by SDSS     Our first ZTF photometry of SN 2023eiw  was obtained on 2023 March 29 (MJD 60032.15) with P48. The first detection was in \textit{g} band, with host-subtracted magnitude of $18.92\pm0.10$mag, at $\alpha$  = $12^\text{h}28^\text{m}46^\text{s}.200$, $\delta$ = $+\ang{46}31'15{''}.64$ (J2000). The Object was first reported to the TNS by \cite{2023eiw} on March 31, and was first detected by \cite{2023eiw} on March 30 at \textit{i}=18.53 mag \citet{2023eiw_detec}. The last ZTF non-detection was on 2023 March 26 at \textit{g} $>$ 20.66 mag. The transient was classified as an SN Type Ic-BL by \citet{2023eiw_class} with spectra taken by the Spectral Energy Distribution Machine (SEDM) attached to Palomar 60-inch telescope (P60). Shown in figure \ref{fig:2} is spectra taken form the Alhambra Faint Object Spectrograph and Camera (ALFOSC) on the Nordic Optical Telescope (NOT). The supernova host-galaxy has a known redshift of \textit{z} = 0.025, measured by SDSS spectroscopy.

\section{SN\,2023\lowercase{zeu} (ZTF18\lowercase{abqtnbk})}
    Our first ZTF photometry of SN 2023zeu (ZTF18abqtnbk) was obtained on 2023 December 9 (MJD 60287.11) with P48. The first detection was in \textit{r} band, with host-subtracted magnitude of $18.82\pm0.10$mag, at $\alpha$  = $01^\text{h}36^\text{m}49^\text{s}.249$, $\delta$ = $+\ang{01}35'05{''}.68$ (J2000). The object was first reported to the TNS by \cite{2023zeu} on December 9, and was first detected by \cite{2023zeu} on December 9 at ${r}=18.82$ mag \citet{2023zeu_detec}. The last ZTF non-detection was on 2023 November 20 at \textit{g} $>$ 19.21 mag. The transient was classified as an SN Type Ic-BL by \citet{2023zeu_class} with spectra taken by SuperNova Integral Field Spectrograph (SNIFS) on the University of Hawaii 88-inch (UH88) telescope. shown in Figure \ref{fig:2}, SN 2023zeu is located within a AGN with host-galaxy measured redshift of ${z} = 0.03$. 

\section{SN 2024\lowercase{rjw} (ZTF24\lowercase{aayimjt)}}
    Our first ZTF photometry of SN 2024rjw (ZTF24aayimjt) was obtained on 2024 August 03 (MJD 60525.30) with P48. The first detection was in the \textit{r} band, with a host-subtracted magnitude of $20.15\pm0.20$mag, at $\alpha$  = $21^\text{h}03^\text{m}10^\text{s}.107$, $\delta$ = $+\ang{20}45'02{''}.58$ (J2000). The object was first reported to the TNS by \cite{2024rjw} on August 5, and was first detected by \cite{2024rjw} on August 3 at \textit{r}=20.15 mag \citet{2024rjw_detec}. The last ZTF non-detection was on 2024 July 31 at \textit{g} $>$ 20.65 mag. The transient was classified as an SN Type Ic-BL by \citet{2024rjw_class} with spectra taken by SuperNova Integral Field Spectrograph (SNIFS) on the University of Hawaii 88-inch (UH88) telescope, shown in figure \ref{fig:2}, and measured redshift of ${z} = 0.02$. 

\section{SN 2024\lowercase{adml} (ZTF24\lowercase{abwsaxu})}
    Our first ZTF photometry of SN 2024adml (ZTF24abwsaxu) was obtained on 2024 December 12 (MJD 60650.46) with P48. The first detection was in the \textit{r} band, with a host-subtracted magnitude of $18.36\pm0.06$mag, at $\alpha$  = $10^\text{h}10^\text{m}40^\text{s}.480$, $\delta$ = $-\ang{02}26'05{''}.14$ (J2000). The Object was first reported to the TNS by \cite{2024adml} on December 6, and was first detected by \cite{2024adml} on December 6 at \textit{r}=18.36 mag by \citet{2024adml_detec}. The last ZTF non-detection was on 2024 December 4 at \textit{g} $>$ 20.10 mag. The transient was classified as an SN type Ic-BL by \citet{2024adml_class} with spectra taken by ESO Faint Object Spectrograph and Camera v2 (EFOSC2) on the New Technology Telescope (NTT, 3.58m). Spectra taken from the Alhambra Faint Object Spectrograph and Camera (ALFOSC) on the Nordic Optical Telescope (NOT) is shown in figure \ref{fig:2}, and has measured redshift of \textit{z} = 0.037.

\section{SN 2024\lowercase{abup} (ZTF24\lowercase{abvtbyt})}
    Our first ZTF photometry of SN 2024abup (ZTF24abvtbyt) was obtained on 2024 December 11 (MJD 60655.17) with the P48. The first detection was in \textit{g} band, with a host-subtracted magnitude of $16.03\pm0.06$mag, at $\alpha$  = $01^\text{h}49^\text{m}11^\text{s}.320$, $\delta$ = $-\ang{10}25'27{''}.44$ (J2000). The object was first reported to the TNS by \cite{2024abup} on November 22, and was first detected by \cite{2024abup} on November 22 at \textit{cyan-ATLAS}=17.02 mag by \citet{2024abup_detec}. The last ATLAS non-detection was on 2024 November 21 at \textit{c} $>$ 19.48 mag. The transient was classified initially as a Type Ib/c by \citet{2024abup_class1} then later as an SN Type Ic-BL by \citet{2024TNSCR4668....1L} with spectra taken by the Wide Field Spectrograph (WiFeS) on the Australian National University (ANU) 2.3m Telescope. The spectra shown in Figure \ref{fig:2} are taken from the Spectral Energy Distribution Machine (SEDM) mounted on the Palomar 60-inch telescope (P60), and has a measured redshift of ${z} = 0.0058$.


\begin{acknowledgements}
T.O-D. and A.C. acknowledge partial support from the NASA \textit{Swift} guest investigator programs (Cycle 18, 19, 20) via awards \#80NSSC23K0314/80NSSC24K1272, \#80NSSC24K1273, and \#80NSSC25K7623, and from the National Science Foundation (NSF) via the award AST-2431072. The National Radio Astronomy Observatory and Green Bank Observatory are facilities of the U.S. National Science Foundation operated under cooperative agreement by Associated Universities, Inc. 
SEDM: SED Machine is based upon work supported by the National Science Foundation under Grant No. 1106171
SA is supported by an LSST-DA Catalyst Fellowship (Grant 62192 from the John Templeton Foundation to LSST-DA). 
S.Y. acknowledges the funding from the National Natural Science Foundation of China under grant No. 12303046, the Startup Research Fund of Henan Academy of Sciences No. 242041217, and the Joint Fund of Henan Province Science and Technology R\&D Program No. 235200810057.
SA also gratefully acknowledges
support from Stanford University, the United States Department of Energy, and a generous grant from Fred
Kavli and The Kavli Foundation. 
B.Z. is supported by Fermi Forward Discovery Group, LLC under Contract No. 89243024CSC000002 with the U.S. Department of Energy, Office of Science, Office of High Energy Physics. This work is based on observations obtained with the Samuel Oschin Telescope 48-inch and the 60-inch Telescope at the Palomar Observatory as part of the Zwicky Transient Facility project. ZTF is supported by the National Science Foundation under Grants No. AST-1440341, AST-2034437, and currently Award \#2407588. ZTF receives additional funding from the ZTF partnership. Current members include Caltech, USA; Caltech/IPAC, USA; University of Maryland, USA; University of California, Berkeley, USA; University of Wisconsin at Milwaukee, USA; Cornell University, USA; Drexel University, USA; University of North Carolina at Chapel Hill, USA; Institute of Science and Technology, Austria; National Central University, Taiwan, and OKC, University of Stockholm, Sweden. Operations are conducted by Caltech's Optical Observatory (COO), Caltech/IPAC, and the University of Washington at Seattle, USA. 
\end{acknowledgements}

\bibliographystyle{aasjournal}
\bibliography{main}

\end{document}